\documentclass[]{misc/imag-ms-template}

\title{\go{ 2.0}: A Next-Generation Toolbox for Brain Segmentation and Cortex Parcellation at Ultra-High Field MRI}

\author{
Marc-Antoine Fortin$^{1\ast}$, Anne Louise Kristoffersen$^{1}$, \\
Kjersti Eline Stige$^{3}$, Nicolas Kunath$^{3,4}$, Charalampos Tzoulis$^{5,6,7}$ \\
and Pål Erik Goa$^{1,2}$\\[0.5em]
\parbox{0.95\textwidth}{\centering\small
$^{1}$ Department of Physics, Norwegian University of Science and Technology, Trondheim, Norway\\
$^{2}$ Department of Radiology and Nuclear Medicine, St. Olavs Hospital HF, Trondheim, Norway\\
$^{3}$ Department of Neurology and Clinical Neurophysiology, St. Olavs Hospital HF, Trondheim, Norway\\
$^{4}$ Department of Neuromedicine and Movement Science, St. Olavs Hospital HF, Trondheim, Norway\\
$^{5}$ Neuro-SysMed Center for Clinical Treatment Research, Department of Neurology, Haukeland University Hospital, 5021 Bergen, Norway\\
$^{6}$ Department of Clinical Medicine, University of Bergen, Pb 7804, 5020 Bergen, Norway\\
$^{7}$ K.G. Jebsen Center for Translational Research in Parkinson’s disease, University of Bergen, Pb 7804, 5020 Bergen, Norway\\
$^{\ast}$ Correspondence: marc.a.fortin@ntnu.no
}
}

\usepackage{comment}
\usepackage{epstopdf}
\usepackage{amsthm}
\usepackage{hyperref}
\usepackage{tabularx}
\usepackage{multirow}
\usepackage{pdflscape}

\newcommand{\go}{GOUHFI}
\newcommand{\ants}{ANTsPyNet}
\newcommand{\nn}{nnU-Net}
\newcommand{\sca}{SCAIFIELD}

\newcommand{\tow}{T1w}

\newcommand{\ttw}{T2w}

\newcommand{\pdw}{PDw}

\newcommand{\fsv}{\textit{FastSurferVINN}}
\newcommand{\fs}{\textit{FastSurfer}}
\newcommand{\fss}{\textit{FreeSurfer}}
\newcommand{\sys}{\textit{SynthSeg$^{+}$}}
\newcommand{\ceb}{\textit{CEREBRUM-7T}}
\newcommand{\spm}{SPM12}

\newcommand{\mpr}{MPRAGE}
\newcommand{\mprr}{MP2RAGE}
\newcommand{\mpl}{MPRAGE$_{like}$}

\newcommand{\sixi}{(0.6 mm)$^{3}$}
\newcommand{\sevi}{(0.7 mm)$^{3}$}
\newcommand{\sevhi}{(0.75 mm)$^{3}$}
\newcommand{\eigi}{(0.8 mm)$^{3}$}
\newcommand{\nini}{(0.9 mm)$^{3}$}
\newcommand{\onei}{1.0 mm$^{3}$}

\begin{document}

\maketitle


\begin{abstract} 
Despite Ultra-High Field MRI (UHF-MRI) being increasingly used in large-scale neuroimaging studies, automatic segmentation and parcellation remain challenging due to signal inhomogeneities, varying contrast and resolution, and the lack of tools optimized for UHF-MRI. Traditional software packages such as \fsv{} or \sys{} often yield suboptimal results when applied directly to UHF images, which has limited region-based quantitative analyses. Building upon this need, we propose \go{ 2.0}, a new implementation of \go{} that incorporates greater training data variation and introduces added functionalities, including cortical parcellation and volumetry.\\
\go{ 2.0} preserves the contrast- and resolution-agnostic properties of the original toolbox while introducing two independently trained segmentation tasks based on the 3D U-Net architecture. The first network segments brain images of any contrast, resolution or field strength into 35 labels, using the domain randomization approach with a dataset composed of 238 subjects of varied resolutions, field strengths and populations. Using the same training dataset, the second network performs the parcellation of the cortex into 62 labels following the Desikan–Killiany–Tourville (DKT) protocol. \\
When evaluated across multiple datasets, \go{ 2.0} demonstrated improved segmentation accuracy relative to the original toolbox, particularly in heterogeneous populations, and its ability to generate reliable cortical parcellations. Additionally, the added integrated volumetry pipeline enabled the derivation of results consistent with those obtained using standard volumetry procedures.\\
In summary, \go{ 2.0} offers a comprehensive, contrast- and resolution-agnostic solution for brain segmentation and parcellation across field strengths. This positions \go{ 2.0} as a versatile tool for researchers working at UHF-MRI, making it the first Deep Learning (DL) toolbox capable of robust cortical parcellation at UHF-MRI. 
\end{abstract}


\section{Introduction}\label{intro}


Historically, the only option that offered both whole-brain segmentation and cortical parcellation was \fss{} for \tow{} images at 1.5 and 3T. With the increased availability and affordability of GPU hardware, several neuroscientists have replaced \fss{} by its DL-equivalent \fs{}, reducing by 25 times (or even more) computation time \citep{henschel_fastsurfer_2020}. Additionally, since sub-millimeter resolution has become increasingly available, a more recent variation of \fs{}, named \fsv{}, was proposed. \fsv{} segments \tow{} images acquired at 1.5 or 3T at native resolution \citep{henschel_fastsurfervinn_2022}. Both \fss{} and \fs{} propose to segment the whole-brain into 35 subcortical regions as originally defined in \cite{fischl2002whole} and the cortex into 62 regions (31 for each hemisphere) following the DKT atlas \citep{klein2012101}.

As a result, over the past two decades, \fss{}, and more recently \fsv{}, have become the gold standard for automatic brain segmentation and cortical parcellation for \tow{} images at 1.5-3T and have been extensively used and validated by the neuroimaging community \citep{chiu2024evaluation,bloch_comparison_2021,khadhraoui2022manual,zughayyar2025fastsurfer}.

Another important neuroimaging toolbox is the Advanced Normalization Tools (ANTs) which has been broadly utilized to study and analyze brain MR images \citep{avants2009advanced}. Lately, a novel implementation of the toolbox employing Deep Learning (DL) has been made available under the names of ANTsRNet/ANTsPyNet corresponding to R and Python implementations of the tool respectively \citep{tustison2021antsx}. While the automatic brain segmentation available through these tools segments the brain into six tissue classes (Cerebrospinal Fluid (CSF), Gray Matter/Cortex (GM), White Matter (WM), deep GM, brainstem and cerebellum) compared to 35 for \fsv{}, the tool \texttt{antspynet.desikan\_killiany\_tourville\_labeling} enables the parcellation of the cortex under the DKT convention, mimicking the \fsv{} ecosystem.

More recently, a novel automatic brain segmentation technique, called \sys{} (also available in the \fss{} ecosystem), has been proposed where a novel DL approach called domain randomization (DR) is used to train a model using synthetic images \citep{billot2023robust}. Following the same labeling convention as defined by \fss{}, \sys{} is able to segment the brain and cortex in a few minutes like \fs{}. The main difference with \fs{} resides in \sys{'s} ability to segment images of any contrast or resolution with a fixed output resolution of \onei{}, allowing the use of non-\tow{} contrasts or lower resolution images as input. \sys{} allows neuroscientists with clinical images acquired with lower image quality than usually acquired in research environments to be used for quantitative analyses like volumetry, which have been impossible previously. Even if proposed recently, the DR approach applied to brain segmentation has been used by several segmentation tools, showing promising avenues for the development of novel techniques for a varied range of applications \citep{valabregue2024comprehensive,gibson2024segcsvdwmh,fortin2025gouhfi}.

While the previously mentioned techniques have been successfully applied and validated, none of them have been widely applied to UHF-MRI (i.e., $\geq$7T). UHF-MRI has been increasingly available in the last decade and used for large neuroimaging studies like the Human Connectome Project (HCP) \citep{van2012hcp}. Despite its several advantages, like higher SNR, contrast and spatial resolution, UHF images typically suffer from significant transmit radiofrequency (RF) inhomogeneities compared to lower fields \citep{trattnig_key_2018}. This results in significant signal and contrast inhomogeneities observed across the image. Although recent developments in parallel transmit (pTx) RF pulses have substantially improved this issue \citep{gras_universal_2017}, their use remains limited and yet to be applied to large neuroimaging studies. 

This inaccessibility to large datasets with homogeneous UHF images has considerably limited the development of DL-based techniques for \tow{} UHF images. Only one technique, \ceb{}, has been designed for \tow{} images acquired with 7T MRI \citep{svanera_cerebrum7t_2021}. However, without fine-tuning or retraining, \ceb{} only segments (0.63mm)$^3$ \tow{} \mprr{} images from the Glasgow dataset with a matrix shape of 256$\times$352$\times$224. Moreover, the images are only segmented into six labels as similarly done by \ants{}. Alternatively, studies have been compelled to use 3T-designed tools like \fss{} on 7T data by implementing extensive preprocessing on the images \citep{zaretskaya2018advantages}. While adapting 3T tools can be a temporary solution for UHF, they do not provide a reliable long-term solution for most studies, especially when they include a substantial number of subjects. Indeed, segmentation results using these tools are frequently unsatisfactory, requiring important visual quality assurance (QA) and extremely time-consuming manual corrections.


Considering this lack of automatic segmentation methods developed for UHF-MRI, we have recently proposed \go{}: Generalized and Optimized segmentation tool for Ultra-High Field Images \citep{fortin2025gouhfi}. By integrating and refining the DR strategy from \sys{} with datasets more tailored to the UHF context for training and testing, \go{} proposes an inhomogeneity-resistant, contrast- and resolution-agnostic brain segmentation tool capable of segmenting the brain into 35 labels following the \fss{} label convention. Considering there is yet to be a standard high-resolution at UHF and many variations of \tow{} protocols are used in the community, the contrast and resolution agnosticity of \go{} revealed itself as a promising tool for segmenting UHF images, especially in cases with significant signal inhomogeneities. \go{} is available online as an open-source toolbox at the following address: \href{https://github.com/mafortin/GOUHFI}{https://github.com/mafortin/GOUHFI}. 

In its original deployment, no cortex parcellation was developed as part of \go{}. This has limited its relevance for UHF neuroimaging studies, particularly functional MRI (fMRI), where this step is frequently performed. Additionally, to the best of our knowledge, no tool is currently capable and designed to robustly perform high quality automatic cortex parcellation following the DKT convention at UHF-MRI. 

While extensively validated on six contrasts and seven resolutions at 3T, 7T and 9.4T, \go{} was mainly validated on healthy and young adults. However, the STRAT-PARK dataset \citep{stige2024strat}, which comprises people with Parkinson's disease (PwP) and age-matched healthy controls (HC), was used to evaluate \go{} in a clinical cohort setting. While \go{} performed well in most subjects, few exposed \go{'s} limitations. More precisely, subjects with substantially enlarged lateral ventricles (also frequently observed for healthy but older brain anatomies), were challenging for \go{}. Indeed, proper delineation of the caudate, thalamus and hippocampus adjacent to large ventricles was sub-optimal, resulting in poor identification of the actual borders between the neighboring regions. While impacting a minority of test cases, this observation raised the need for a more exhaustive validation on non-healthy and older brain anatomies for \go{}. 

With the increasing adoption of UHF-MRI in the neuroimaging community for large-scale studies like the HCP, the lack of robust automatic cortical parcellation and the limitations of \go{} in segmenting aged brains underscore the need for further development and improvement of the toolbox.

In this work, we present \go{ 2.0}, a novel, modified and improved version of the originally proposed tool, where automatic brain segmentation and cortex parcellation are performed with two trained DL models using a training corpus with increased anatomy variety including aged and demented subjects. Additionally, \go{ 2.0} integrates a volumetry pipeline estimating volumes of individual structures as well as total intracranial volume (TIV), thereby providing a complete solution for segmentation and volumetric analyses without relying on external softwares.  

In summary, we first demonstrate how \go{ 2.0} was developed in contrast to the original \go{}. We then present analyses assessing: (1) the contrast-agnostic performance of \go{ 2.0} at UHF compared to its predecessor, (2) its robustness in clinical cohorts of Parkinson's Disease and SpinoCerebellar Ataxias (SCA) patients at 7T, (3) the accuracy of its TIV estimations against gold-standard and competing techniques, and (4) its cortical parcellation accuracy against manual delineations at 3T, benchmarked against existing tools. Finally, we provide a discussion of these analyses, together with the current limitations of \go{ 2.0} and directions for future work.

\section{Methods}\label{meth}

\subsection{Datasets}

Most of the data used in this study are the same as in \cite{fortin2025gouhfi}. An overview of all these datasets is provided in Table \ref{tab:data} below, where the training and test datasets are described. In this study, three datasets were added, one for training and two for testing, and are presented in the following subsections.

\begin{table}[h]
    \scriptsize
    \centering
    \caption{Summary of the datasets used for training and testing in this work. The new datasets are in bold. The table lists the field strength, resolution, contrast, subject type, vendor, and number of subjects. ASD: Autism Spectrum Disorder, AD: Alzheimer's Disease, PwP: People with Parkinson's Disease, SCA: SpinoCerebellar Ataxia.}
    \begin{tabular}{l r r r r r r}
        \toprule
        \multicolumn{7}{c}{\textbf{Training Datasets}} \\
        \midrule
        Dataset & Field Strength & Resolution & Contrast & Subjects & Vendor & N \\
        \midrule
        HCP-YA & 3T & \sevi{} & \tow{}/\ttw{} & Healthy & Siemens & 80 \\
        SCAIFIELD & 7T (pTx) & \sixi{} & \tow{}, \mpl{}, MPM-T1w,-MTw,-PDw & Healthy & Siemens & 31 \\
        UltraCortex & 9.4T (1Tx) & \sixi{} & \tow{} & Healthy & Siemens & 15 \\
        ABIDE-II ETHZ & 3T & \nini{} & \tow{} & ASD & Philips & 34 \\
        ABIDE-II EMC & 3T & \nini{} & \tow{} & ASD & GE & 46 \\
        \textbf{OASIS2} & 1.5T & 1 mm$^3$ & \tow{} & Healthy \& AD & Siemens & 32 \\
        \midrule
        \multicolumn{7}{c}{\textbf{Test Datasets}} \\
        \midrule
        Dataset & Field Strength & Resolution & Contrast & Subjects & Vendor & N \\
        \midrule
        \textbf{OASIS-TRT-20} & 1.5T & 1 mm$^3$ & \tow{} & Healthy & Siemens & 20 \\
        HCP-YA & 3T & \sevi{} & \tow{}/\ttw{} & Healthy & Siemens & 20 \\
        SCAIFIELD & 7T (pTx) & \sixi{} & \tow{}, \mpl{}, MPM-T1w,-MTw,-PDw & Healthy & Siemens & 10 \\
        \textbf{SCAIFIELD-NPC} & 7T (pTx) & \sixi{} & \tow{}, \mpl{}, MPM-T1w,-MTw,-PDw & SCA & Siemens & 3 \\
        STRAT-PARK & 7T (1Tx) & \sevhi{} & \tow{} & PwP/Healthy & Siemens & 45 \\
        \bottomrule
    \end{tabular}
    \label{tab:data}
\end{table}

\subsubsection{Open Access Series of Imaging Studies:  OASIS2 \& OASIS1-Test-Retest}

A subset of 32 subjects from the Open Access Series of Imaging Studies 2 (OASIS2) \citep{marcus2007open} was selected in order to increase the anatomical variation in the training dataset by adding demented and non-demented subjects aged between 72 and 97 years old. For each subject, a \onei{} \tow{} MPRAGE was acquired with a 1.5T Vision scanner (Siemens Healthineers, Erlangen, Germany). Subjects were selected based on the presence of substantially enlarged ventricles, creating anatomical variations impossible to create through the generation of synthetic images with the generative model. The complete OASIS2 dataset and information about the dataset can be accessed online\footnote{\url{https://sites.wustl.edu/oasisbrains/home/oasis-2/}}.

A second subset of the OASIS data was utilized for testing \go{ 2.0}. The manually delineated cortical parcellation done on 20 subjects from the OASIS1-Test-Retest (OASIS-TRT-20) dataset, also known as the MindBoggle-101 dataset \citep{klein2012101}, was used. The manual delineations were made from a \onei{} \tow{} MPRAGE and follow the DKT protocol (31 cortical labels per hemisphere), which is the same protocol used by \fsv{}. To the best of our knowledge, this is the only set of manually segmented cortex parcellations following this protocol available online\footnote{\url{https://mindboggle.info/data.html}}.

\subsubsection{SCAIFIELD-NPC: Spinocerebellar Ataxia - Norwegian Patient Cohort}

In this study, we used the SCAIFIELD Norwegian Patient Cohort (SCAIFIELD-NPC) as a test dataset. While the imaging protocol is exactly the same as for the SCAIFIELD protocol, the difference resides in the fact that it only includes three SCA subjects instead of healthy controls. SCA is a heterogeneous group of neurodegenerative genetic disorders characterized by progressive cerebellar and brain atrophy \cite{klockgether2011update}. Because SCA induces substantial brain atrophy, this dataset was used strictly for a qualitative assessment of the segmentation outputs from \fsv{}, \sys{}, and \go{2.0}, rather than for quantitative performance comparison, in a 7T clinical cohort.

Each study was approved by the local review boards of each site/institution and participants of the individual studies signed a written informed consent form before scanning. Complete ethic statements are available at each respective study web pages and publications.

In total, the training corpus was composed of 238 different subjects from the HCP (n=80), SCAIFIELD (n=31), UltraCortex (n=15), ABIDE-II (n=80) and OASIS2 (n=32) datasets.

\subsection{Data processing}

Except for the modifications mentioned below, the rest of the processing of the training data was kept untouched from that described in the original \go{} publication and refer the reader to it \citep{fortin2025gouhfi}. 

\subsubsection{Initial training label maps creation with \fsv{}}

All \tow{} images used in this study were segmented using \fsv{} \citep{henschel_fastsurfervinn_2022} (v2.3.0) with the \textit{--seg\_only} flag in order to produce automatic whole brain segmentations into 95 labels (33 subcortical labels and 62 cortex labels following the DKT labeling protocol). The list of labels produced by \fsv{}, and used in this work, is available in Table 5 of Appendix A of \cite{henschel_fastsurfervinn_2022}.

\subsubsection{Creation of new training label maps for brain segmentation and cortex parcellation for \go{ 2.0}}

\begin{figure}[htbp]
    \begin{center}
        \includegraphics[width=0.99\textwidth]{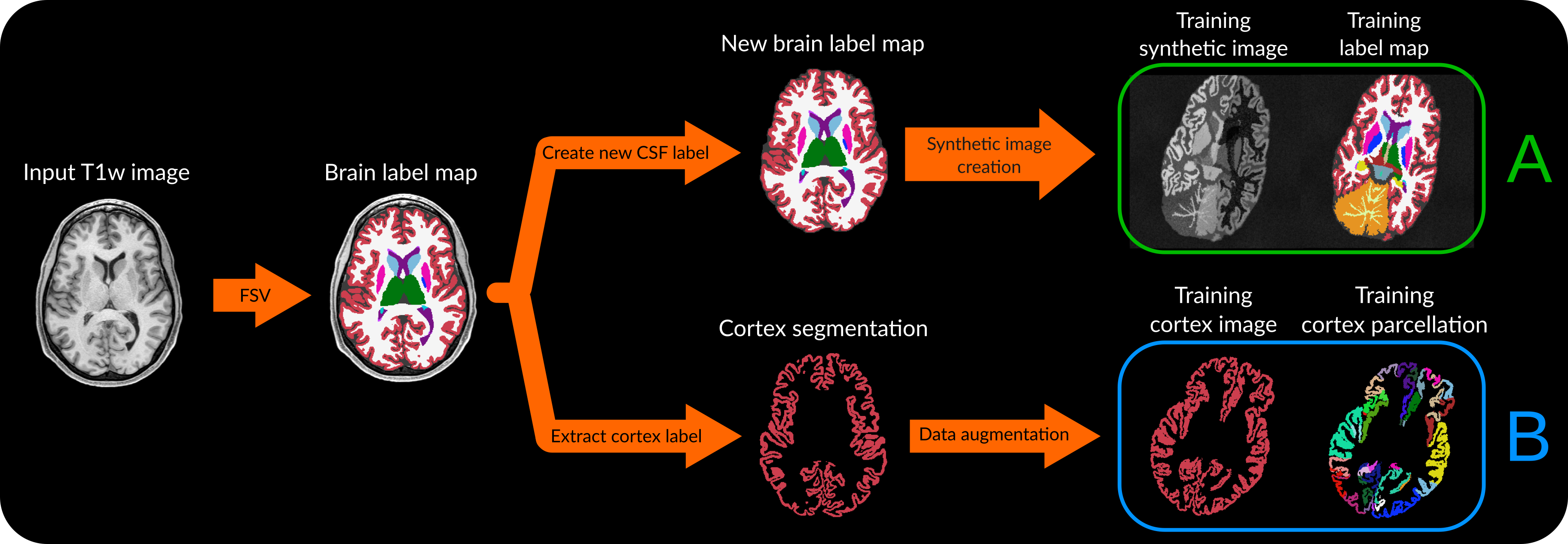}
        \caption{Pipeline to create the training data for both 3D U-Nets used in \go{ 2.0}. For all training cases, the \tow{} image is segmented using \fsv{} (FSV) at native image resolution. The label map is then used for two things. On one side, the CSF label is modified to include all unassigned brain voxels present inside the brain mask. This new label map is then fed into the generative model to create the domain randomized synthetic image and corresponding label map. This enables to create a dataset of synthetic images (A) for training a contrast agnostic 3D U-Net for brain segmentation. Separately, the cortex segmentation is extracted from the label map by masking all remaining labels. This cortex segmentation, with its corresponding parcellation, are fed into the same domain randomization model as (A) without incorporating signal inhomogeneity since these are label maps and not images. This results in a second training dataset (B) used to train a 3D U-Net to perform cortex parcellation from an input cortex segmentation.}
        \label{meth:data}
    \end{center}
\end{figure}

In comparison to the original \go{}, no "Extra-Cerebral" label was added to the training label maps for brain segmentation for \go{ 2.0}. Instead, the original CSF label was modified by adding all unassigned voxels inside the brain segmentation mask provided by \fsv{} as CSF (i.e., all voxels not assigned to any label inside the intracranial space were then assigned to CSF). Moreover, no dilation was performed as originally done for the original \go{}. Regarding the cortex parcellation, the cortex segmentation (one cortex label for each hemisphere, no subdivision) and its corresponding parcellation (cortex subdivision into 62 labels) were extracted from the \fsv{} label map. Both training data creation processes are shown in Figure \ref{meth:data}.

\subsubsection{Training datasets: Brain segmentation and cortex parcellation}

\begin{figure}[htbp]
    \begin{center}
        \includegraphics[width=0.95\textwidth]{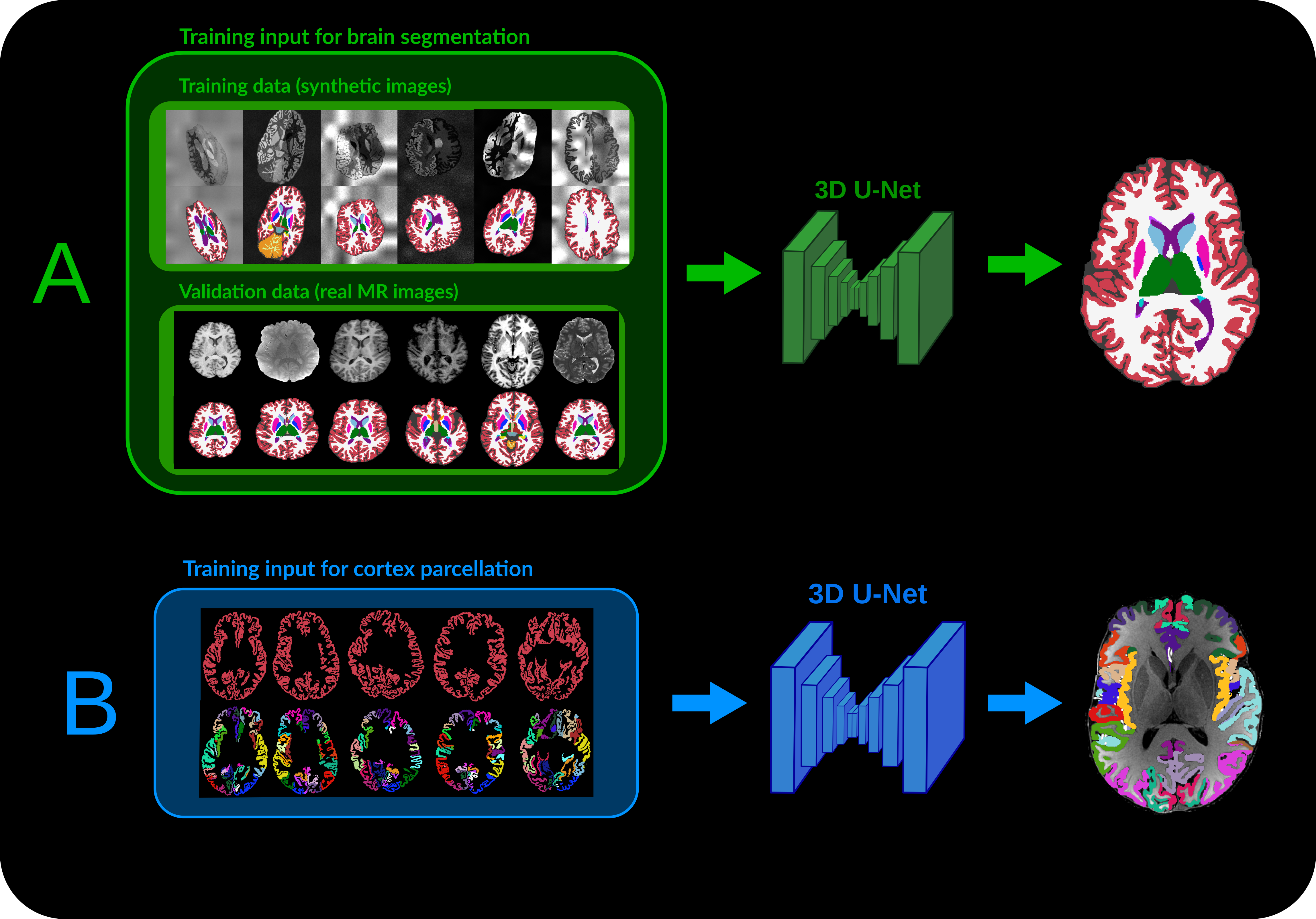}
        \caption{Training datasets used for both 3D U-Nets trained as part of \go{ 2.0}. Dataset A corresponds to the training dataset used for the contrast-agnostic brain segmentation task as done for the original \go{}. One important detail for dataset A is that the training data is composed of synthetic images whereas the validation data is composed of real MR images in order to reflect the true usage domain with real MR images. Alternatively, dataset B consists only of discrete label maps where a single cortex segmentation label is used as training "image" and its corresponding cortex parcellation with 62 labels is the training label map.}
        \label{meth:datasets}
    \end{center}
\end{figure}

In contrast with its original implementation, \go{ 2.0} now utilizes two 3D U-Nets \citep{ronneberger2015u} where one network is used for the brain segmentation task and the other one for the cortex parcellation. Regarding the first network (A in Figure \ref{meth:datasets}), the training label maps were used to generate synthetic images following the domain randomization (DR) approach proposed in \cite{billot2023synthseg} adapted to the UHF-MRI context as explained in \cite{fortin2025gouhfi}. While the training corpus consisted of synthetic images, the validation set was composed of the corresponding real masked \tow{} image for each subject, except for the HCP-YA and \sca{}, where the coregistered \ttw{} and \pdw{} images were used. Validating the model on real MR images allows to check if the model is able to learn domain-agnostic features that translates well into the actual intended usage with real MR images and avoids learning features specific to synthetic images only. 

Moreover, \cite{henschel_fastsurfervinn_2022} showed that enlarging the training corpus was the key factor for improving both performance and generalizability of DL segmentation models. Motivated by this, we designed two training dataset variants for the brain segmentation task in \go{ 2.0}. The first variant, \go{ 2.0-n1}, followed the original \go{} strategy, where one synthetic image was generated per input label map for the generative model (238 synthetic images). Since the generation procedure synthesizes images with widely varying contrasts, artifacts, and anatomical shapes, a second variant where two synthetic images per input label map was tested. This second variant, named \go{ 2.0-n2}, had a total of 476 synthetic images in its training corpus. 

For the second U-Net, (B in \ref{meth:datasets}), the training corpus was composed of label maps with only the cortex label as training data. Indeed, this network was not trained to segment images, but to subdivide the cortex label into 62 labels (or parcellations). No synthetic image was used for this model, but the same random spatial augmentation process was used to generate the unrealistically augmented training cortex labels. In other words, the same DR approach was used except that the random generation of signal inhomogeneity was disabled since the training inputs were label maps with discrete values and not images. For this training dataset, the 238 augmented cortex labels were used.

A schematic of both training corpus can be seen in Figure \ref{meth:datasets}

\subsection{Deep Learning Framework of \go{ 2.0}}

\subsubsection{Network architecture \& Training setup}

The \nn{} framework (v2.4.1) was used to implement the two 3D U-Nets with residual encoders \citep{isensee2021nnu,isensee2024nnu}. The first U-Net doing the brain segmentation into 35 labels was composed of six layers with 32, 64, 128, 256, 320 and 320 features, using Leaky ReLU \citep{maas2013rectifier}, 3$\times{}$3$\times{}$3 kernel size, patch size of 224$\times$160$\times$192 and a batch size of 2. The loss function was the sum of the soft Dice and cross-entropy losses \citep{drozdzal2016importance}. The median resolution of the training dataset was set to \eigi{}, which was then used as the training resolution for both U-Nets. For the first model using synthetic training images, 3D cubic spline and linear interpolation with one-hot encoding were used for resampling the training images and label maps respectively. For the second model using only the cortex label maps as training images, no normalization on the input label maps was done  with 3D linear interpolation with one-hot encoded for resampling both the training label maps and corresponding cortex parcellation label maps to the median training resolution. A more detailed description of all steps performed by the \nn{} framework can be found in \cite{isensee2021nnu}.

Both U-Nets were trained following identical hyper-parameters. Using 5-fold cross-validation, the 238 subjects were randomly separated in five training datasets with 80\% of the subjects assigned as training and the remaining 20\% as validation. For each fold, 500 epochs were used, where one epoch was defined as 250 random mini-batches fed to the network. An early stop criterion set to 60 consecutive epochs without improvement to the validation pseudo dice score. The AdamW optimizer \citep{loshchilov2017decoupled} was utilized with a base learning rate (LR) of 3$\times$10$^{-4}$ decaying following the poly LR scheduler. The training lasted approximately three days and was done using an NVIDIA Ampere A40 GPU with 48 Gb of VRAM.

\subsubsection{Inference}

Since the cortex parcellation model (the second U-Net) uses the cortex segmentation generated by the first brain segmentation model, the second model needs to be ran after the first. In other words, the user can only do brain segmentation alone without executing the cortex parcellation, but not the other way around.

\begin{figure}[htbp]
    \begin{center}
        \includegraphics[width=0.95\textwidth]{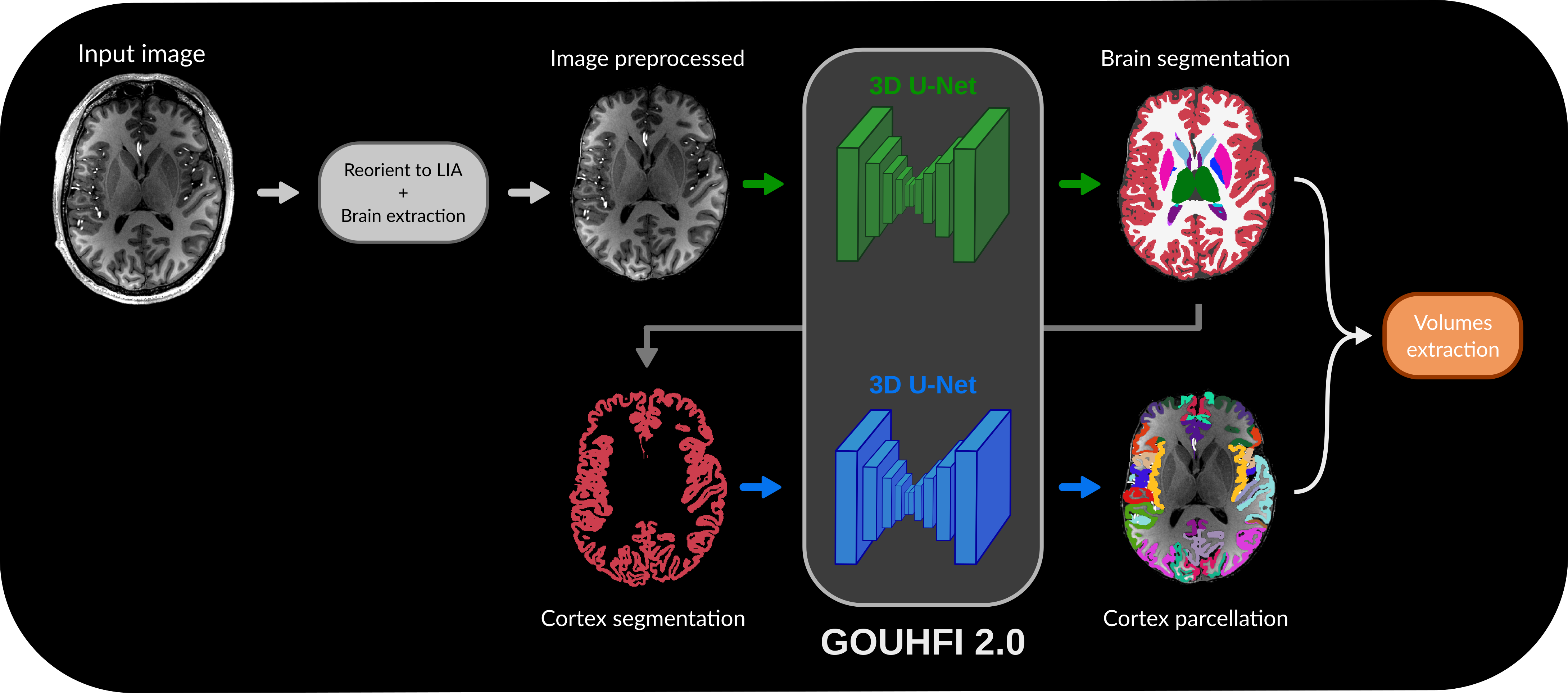}
        \caption{\go{ 2.0} processing pipeline. By using an image of any contrast, resolution and even field strength, one can obtain (1) a whole brain segmentation into 35 labels following the \fss{} lookup table and (2) cortex parcellation into 62 labels (31 per hemisphere) following the Desikan-Killiany-Tourville (DKT) convention. The volume of every structure in the output brain segmentation and cortex parcellation are calculated and exported for subsequent quantitative analyses. While not developed as part of \go{ 2.0}, the reorientation and brain extraction steps are implemented in the toolbox, making it a completely standalone solution for brain neuroimaging analyses like \fss{}/\fsv{} or \sys{}.}
        \label{meth:pipe}
    \end{center}
\end{figure}

Since 5-fold cross-validation was used, an ensembling strategy where the softmax outputs from all five models are averaged together was used to produce a single "hard" label map for each U-Net. Inference, post-processing and resampling (if using a different resolution from the training one) took approximately 60 seconds per case for both models together in total. Figure \ref{meth:pipe} shows the overall pipeline proposed with \go{ 2.0}.

\subsection{Evaluation}

\subsubsection{Competing methods}

For brain segmentation, \fsv{} (v2.3.0) \citep{henschel_fastsurfervinn_2022}, \sys{} \citep{billot2023robust} and the original \go{} were used for qualitative comparison with the two variants of \go{ 2.0}. For the quantitative comparison, \fsv{} was excluded due to its limitations at UHF and inability to segment non-\tow{} contrasts. While all above-mentioned methods segment at native input image resolution, \sys{} generates segmentation outputs at a standardized resolution of \onei{}, independent of the input resolution. Consequently, to obtain label maps at native resolution and allow quantitative comparisons, the same external up-sampling strategy using one-hot encoded 3D linear interpolation for label maps as done for \go{} was implemented in-house for \sys{'s} label maps. For cortex parcellation, the mentioned methods were used, in addition to \ants{} \citep{tustison2021antsx}. 

\subsubsection{Quantitative evaluation}

In order to assess the quality of the segmentations produced by \go{ 2.0}, the Dice-Sørensen Similarity Coefficient (DSC) \citep{dice1945measures,sorensen1948method}, which measures the overlap between two segments (with a value of 1 being a perfect overlap between the two segments) was computed with the following equation: 

\begin{equation}
    DSC = \frac{2 \times |G \cap P|}{G + P}
\end{equation}

where G is the ground truth segment, and P is the predicted segment to be compared.

Moreover, the Average Surface Distance (ASD) \citep{reinke2024understanding}, where a value of 0 represents a perfect alignment of both surfaces evaluated, was computed with the following equation:  

\begin{equation}
    ASD = \frac{\displaystyle\sum_{i=1}^{N_G} d_{G\rightarrow P,i} + \sum_{i=1}^{N_P} d_{P\rightarrow G,i}}{N_G + N_P}\quad .
\end{equation}

Herein, $d_{G\rightarrow P,i}$ is the distance from point $i$ on the surface of the ground truth segment to its nearest point on the surface of the predicted segment; $d_{P\rightarrow G,i}$ is the distance from point $i$ on the surface of the predicted segment to its nearest point on the surface of the ground truth segment; $N_G$ and $N_P$ are the total number of points on the ground truth and predicted surfaces respectively.

For calculating DSC and ASD, the choroid plexus and WM-hypointensities were excluded since \sys{} does not segment these labels. Consequently, the lateral and inferior lateral ventricles (both hemispheres) were also excluded since both regions are directly impacted by the presence of the choroid plexus label. Finally, the CSF was also excluded since \sys{} and \go{ 2.0} use different definitions of CSF. In case a label was missing in a label map (i.e., not segmented), DSC and ASD values were set to 0 and NaN respectively. 

For the cortex parcellation, 52 cortical labels were used out of the 62 segmented in order to include \sys{} in the quantitative assessment. Indeed, \sys{} follows the labeling convention used by \fss{} with 68 cortical structures, whereas \go{ 2.0}, \fsv{} and \ants{} follow the DKT labeling protocol with 62 cortical labels. From an ad hoc inspection of both labeling conventions, 10 labels had to be excluded in order to avoid negatively impacting the overlap with the ground truth.

\subsubsection{Total intracranial volume estimation and volumetry analysis}

The volume of individual brain structures was computed by summing all voxels present in the final label maps and multiplying it by the voxel size. For the total intracranial volume (TIV), all structures were summed together. In order to assess the TIV estimation from both \go{ 2.0} variants, the computed TIV values were compared with the estimation from \sys{} and SPM12 \citep{ashburner2012spm}, the latter being considered the gold standard technique \citep{malone2015accurate}.

As part of the assessment of modifying the training corpus for \go{ 2.0}, the same volumetric study as performed for the original \go{} was done on the STRAT-PARK clinical cohort with PwP at 7T. For both HC and PwP, the mean group volume for the putamen, amygdala and hippocampus, normalized by TIV, were computed based on the segmentations produced by \fsv{} (reference), \go{ 2.0} and \sys{}. These ROIs were selected based on the literature for PD \citep{junque2005amygdalar,pieperhoff2022regional,geng2006magnetic}. A Mann-Whitney U test \citep{mann1947test} with Bonferroni-corrected p-values \citep{bonferroni1936teoria} was computed to measure the statistical differences between each group.

\section{Results}\label{res}

\subsection{\go{ 2.0}: Addressing challenges in older and clinical cohorts} \label{res:why2p0}

\begin{figure}[htbp]
    \begin{center}
        \includegraphics[width=0.69\textwidth]{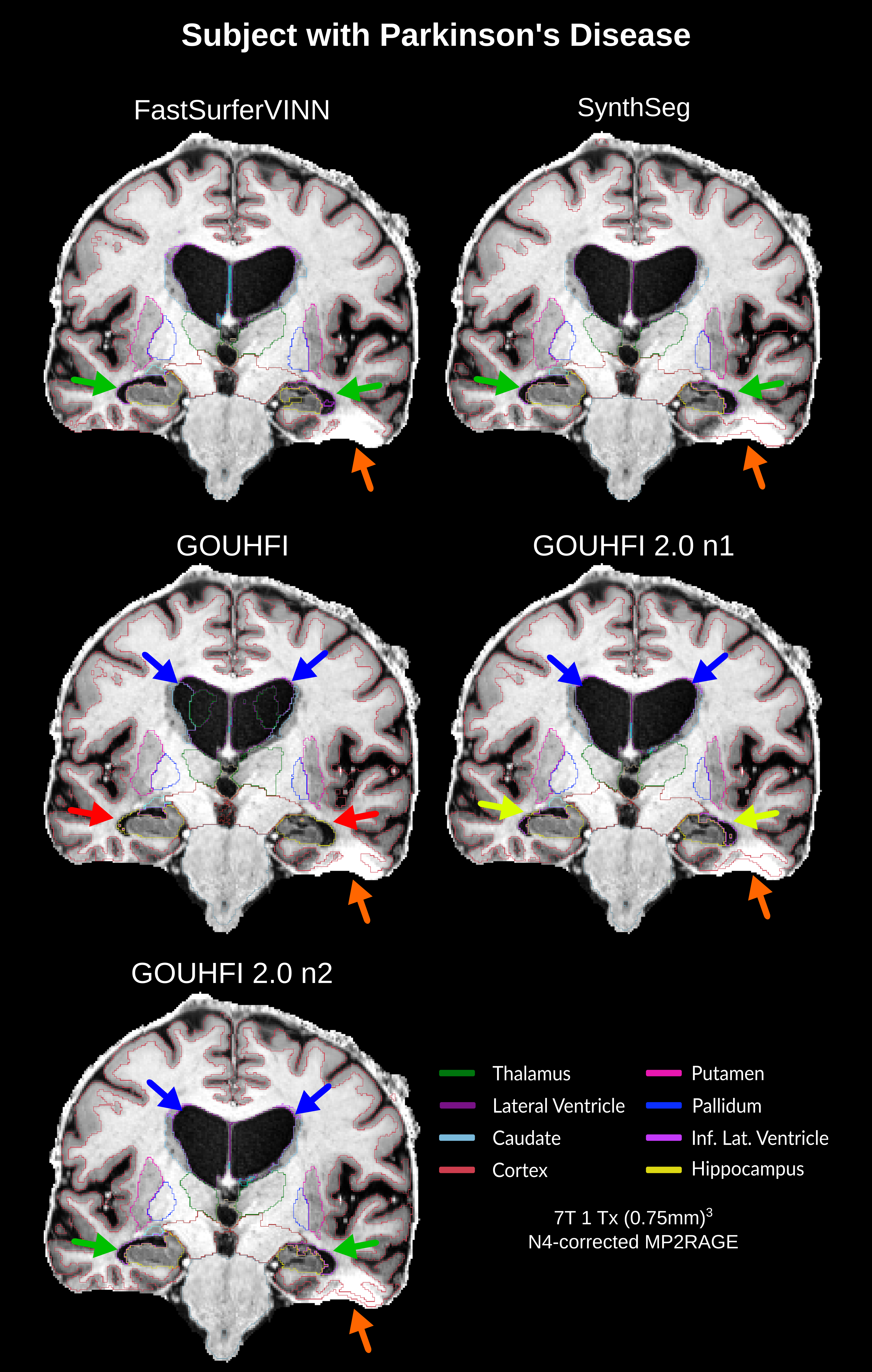}
        \caption{Figure presenting the differences in segmentations between \fsv{}, \sys{}, \go{}, \go{ 2.0-n1} and \go{ 2.0-n2}, computed from an N4-corrected \mprr{} image highly affected by signal inhomogeneities. The example is from a PwP subject in the STRAT-PARK dataset, exhibiting substantial ventricular enlargement. Red, yellow and green arrows point to the quality of the hippocampus-inferior lateral ventricle delineations across techniques. Blue arrows demonstrate the repeated improvement of the segmentation of enlarged ventricles throughout \go{'s} versions. Orange arrows demonstrate the performance of the different algorithms regarding the segmentation of the cortex in presence of highly inhomogeneous signal.}
        \label{fig:why-2p0}
    \end{center}
\end{figure}

Figure \ref{fig:why-2p0} shows the performance of all segmentation techniques tested for a PwP subject acquired at 1 Tx 7T \mprr{}. In cases where subjects would exhibit enlarged ventricles either due to pathology or age, \go{} showed two main limitations which were specific to this approach. First, in a single subject of the 45 in the STRAT-PARK dataset, \go{} segmented the caudate and some parts of the thalamus inside the lateral ventricles (blue arrows in Figure \ref{fig:why-2p0}). It is important to mention that this behavior was only observed in one subject in the STRAT-PARK dataset (out of 45). Second, \go{} often resulted in poor delineation of the boundary between the hippocampus and inferior lateral ventricles (red/yellow/green arrows on Figure \ref{fig:why-2p0}). None of the two variants of \go{ 2.0} exhibited these limitations. However, a slightly improved hippocampus segmentation and inhomogeneity resistance (orange arrow) was observed for \go{ 2.0-n2} compared to \go{ 2.0-n1}. 

\subsection{\sca{}: Comparing \go{'s} variants in healthy controls at 7T}\label{res:sca}

\begin{figure}[htbp]
    \begin{center}
        \includegraphics[width=0.9\textwidth]{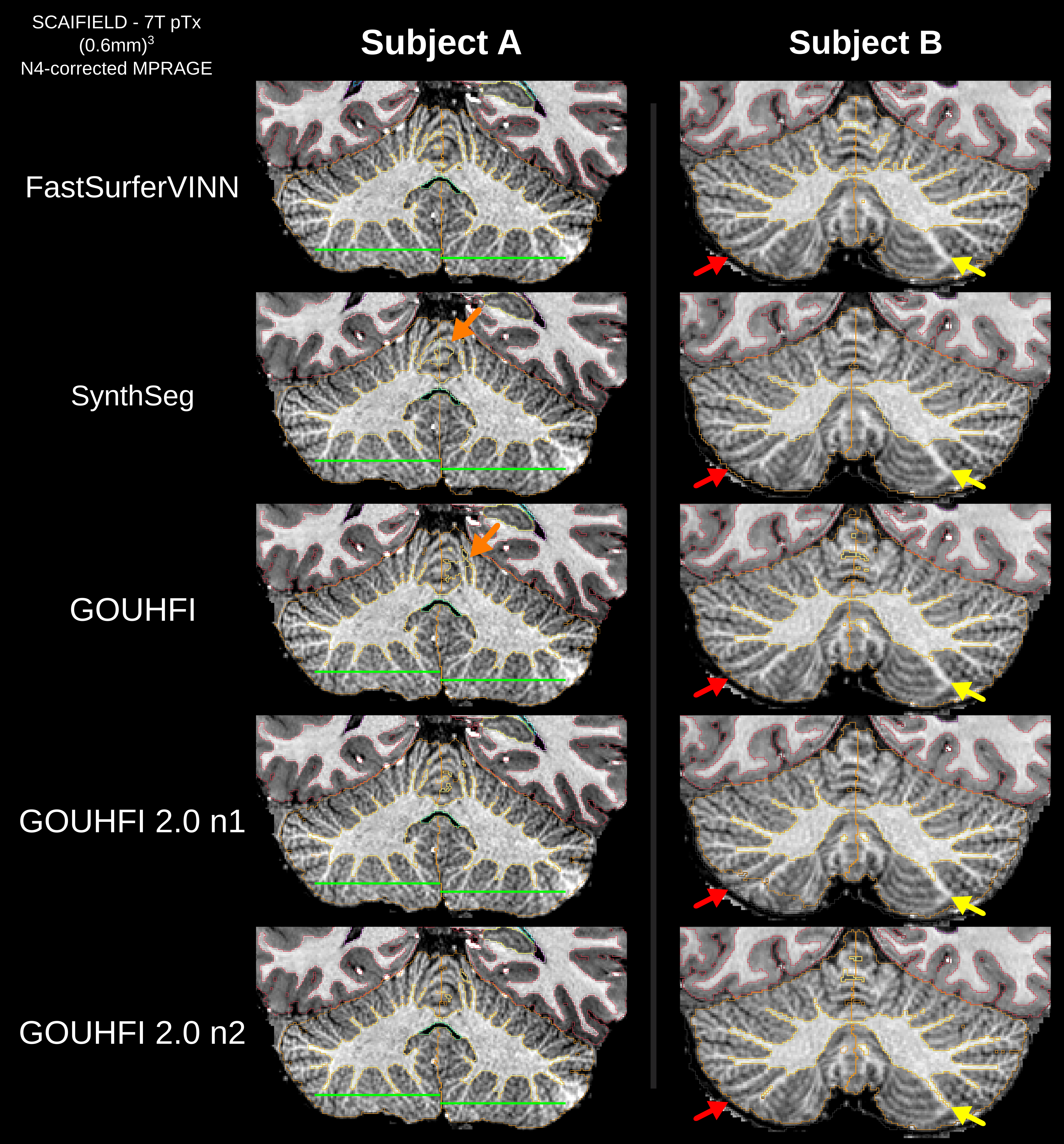}
        \caption{Figure demonstrating the difference in cerebellum segmentations between \fsv{}, \sys{}, \go{}, \go{ 2.0-n1} and \go{ 2.0-n2} computed from the N4-corrected \mpr{} image for two healthy subjects in the SCAIFIELD dataset. For subject A, the green horizontal lines correspond to the most inferior posterior cerebellar WM branches detected by \go{ 2.0-n2} for both hemispheres. The lines for \go{ 2.0-n2} are reproduced for every method for comparison. The orange arrows show errors in cerebellar WM segmentation. For subject B, red arrows show the mislabeling of the cerebellar cortex by \go{ 2.0-n1} compared to the other techniques. Yellow arrows compare the segmentation of one cerebellar WM branch between techniques.}
        \label{fig:why-2p0n2}
    \end{center}
\end{figure}

A coronal slice of an N4-\mpr{} image from two healthy subjects in the \sca{} dataset is shown in Figure \ref{fig:why-2p0n2} for all segmentation techniques tested in this work. For subject A, the delineation of cerebellum WM and cortex is shown, with a special focus on the posterior cerebellum WM branches. 

All \go{} variants demonstrated an improved identification compared to \fsv{} and \sys{}, with a slight but detectable improvement for \go{ 2.0-n2}. Moreover, \go{} and \sys{} over-detection of superior cerebellum WM (orange arrows) appeared to be substantially reduced with both \go{ 2.0} variants. For subject B, although the posterior cerebellar white matter branches are still visible (yellow arrows), particular attention is drawn to the delineation of the posterior cerebellar cortex (red arrows). The \go{ 2.0-n1} variant was the only technique that repeatedly faced problems to properly delineate the posterior cerebellum cortex. That issue was not present in the original \go{}. Overall, \go{ 2.0-n2} demonstrated the best cerebellum WM branches identification while not being affected by the suboptimal cerebellum cortex segmentation, as was the case for \go{ 2.0-n1}.

In order to quantitatively assess the differences in segmentation between the original \go{} and the two new variants, the median and interquartile range (IQR) for the DSC and ASD for several brain regions for the N4-\mpr{} and \mpl{} \citep{fortin2025} were computed, and are listed in Table \ref{tab:sca-dsc-asd-mp}. The two variants of \go{ 2.0} showed highly similar performance, markedly outperforming \sys{} for all labels with a difference of 6 DSC points on average and close to double the ASD values. \go{ 2.0-n2} has the highest DSC and ASD values compared to \go{}.

\begin{table}[h]
\centering
\begin{threeparttable}
\caption{Median DSC and ASD values (with IQR) computed for the N4-\mpr{} and \mpl{} in the SCAIFIELD dataset (n=10) using \go{ 2.0-n1}, \go{ 2.0-n2} and \sys{}. The ground truth is the segmentation created by \go{}. The highest DSC (and lowest ASD) values among the three methods for each structure are shown in bold.}
\scriptsize
\begin{tabular*}{\linewidth}{@{\extracolsep{\fill}}l ccc @{\hskip 3em} ccc}
\toprule
\textbf{DSC} & \multicolumn{3}{@{\hskip 1em}c@{\hskip 1em}}{N4-\mpr{}} & \multicolumn{3}{@{\hskip 1em}c@{\hskip 1em}}{\mpl{}} \\
            & \go{ 2.0-n1} & \go{ 2.0-n2} & \sys{} & \go{ 2.0-n1} & \go{ 2.0-n2} & \sys{} \\
\midrule
WM                & \textbf{0.98 [0.98, 0.98]} & \textbf{0.98 [0.98, 0.98]} & 0.93 [0.92, 0.93] & 0.98 [0.98, 0.98] & \textbf{0.99 [0.98, 0.99]} & 0.92 [0.92, 0.93] \\
\addlinespace
Cortex            & \textbf{0.95 [0.95, 0.95]} & \textbf{0.95 [0.95, 0.95]} & 0.88 [0.87, 0.88] & 0.95 [0.95, 0.96] & \textbf{0.96 [0.96, 0.97]} & 0.87 [0.85, 0.88] \\
\addlinespace
Putamen           & \textbf{0.97 [0.97, 0.97]} & \textbf{0.97 [0.97, 0.97]} & 0.92 [0.92, 0.92] & \textbf{0.97 [0.96, 0.97]} & \textbf{0.97 [0.97, 0.97]} & 0.91 [0.90, 0.92] \\
\addlinespace
Thalamus          & \textbf{0.96 [0.95, 0.97]} & \textbf{0.96 [0.96, 0.97]} & 0.92 [0.91, 0.93] & \textbf{0.97 [0.96, 0.97]} & \textbf{0.97 [0.96, 0.98]} & 0.93 [0.92, 0.94] \\
\addlinespace
Pallidum          & 0.93 [0.92, 0.95] & \textbf{0.95 [0.94, 0.95]} & 0.88 [0.86, 0.88] & 0.93 [0.92, 0.95] & \textbf{0.95 [0.94, 0.96]} & 0.88 [0.86, 0.89] \\
\addlinespace
Cerebellum WM     & \textbf{0.94 [0.94, 0.95]} & \textbf{0.94 [0.94, 0.94]} & 0.88 [0.87, 0.89] & 0.94 [0.94, 0.95] & \textbf{0.96 [0.95, 0.96]} & 0.88 [0.88, 0.89] \\
\addlinespace
Cerebellum Cortex & 0.94 [0.94, 0.95] & \textbf{0.95 [0.95, 0.96]} & 0.93 [0.91, 0.93] & \textbf{0.96 [0.95, 0.96]} & \textbf{0.96 [0.95, 0.96]} & 0.91 [0.90, 0.91] \\
\addlinespace
\textbf{Average (27 labels)} & \textbf{0.95 [0.93, 0.96]} & \textbf{0.95 [0.93, 0.96]} & 0.89 [0.87, 0.93] & 0.95 [0.93, 0.96] & \textbf{0.96 [0.94, 0.97]} & 0.89 [0.86, 0.92] \\
\bottomrule
\end{tabular*}
\vspace{1em}
\begin{tabular*}{\linewidth}{@{\extracolsep{\fill}}l ccc @{\hskip 3em} ccc}
\toprule
\textbf{ASD [mm]} & \multicolumn{3}{@{\hskip 1em}c@{\hskip 1em}}{N4-\mpr{}} & \multicolumn{3}{@{\hskip 1em}c@{\hskip 1em}}{\mpl{}} \\

                  & \go{ 2.0-n1} & \go{ 2.0-n2} & \sys{} & \go{ 2.0-n1} & \go{ 2.0-n2} & \sys{} \\
\midrule
WM                & \textbf{0.11 [0.10, 0.11]} & \textbf{0.11 [0.11, 0.12]} & 0.34 [0.33, 0.35] & 0.11 [0.11, 0.12] & \textbf{0.08 [0.07, 0.09]} & 0.34 [0.33, 0.36] \\
\addlinespace
Cortex            & \textbf{0.16 [0.15, 0.16]} & 0.17 [0.15, 0.18] & 0.36 [0.36, 0.37] & 0.15 [0.14, 0.15] & \textbf{0.13 [0.12, 0.13]} & 0.38 [0.38, 0.39] \\
\addlinespace
Putamen           & \textbf{0.17 [0.15, 0.18]} & \textbf{0.17 [0.16, 0.18]} & 0.44 [0.43, 0.46] & 0.17 [0.16, 0.19] & \textbf{0.15 [0.14, 0.17]} & 0.48 [0.43, 0.53] \\
\addlinespace
Thalamus          & \textbf{0.26 [0.23, 0.32]} & 0.27 [0.22, 0.29] & 0.49 [0.46, 0.56] & 0.25 [0.23, 0.28] & \textbf{0.21 [0.18, 0.25]} & 0.45 [0.41, 0.51] \\
\addlinespace
Pallidum          & 0.30 [0.23, 0.35] & \textbf{0.25 [0.19, 0.27]} & 0.50 [0.45, 0.56] & 0.32 [0.23, 0.34] & \textbf{0.22 [0.17, 0.26]} & 0.49 [0.41, 0.57] \\
\addlinespace
Cerebellum WM     & \textbf{0.25 [0.25, 0.28]} & 0.33 [0.29, 0.40] & 0.57 [0.52, 0.65] & \textbf{0.22 [0.20, 0.23]} & 0.24 [0.20, 0.30] & 0.49 [0.46, 0.52] \\
\addlinespace
Cerebellum Cortex & 0.39 [0.37, 0.42] & \textbf{0.38 [0.35, 0.42]} & 0.52 [0.51, 0.57] & \textbf{0.33 [0.30, 0.36]} & 0.36 [0.34, 0.39] & 0.66 [0.61, 0.70] \\
\addlinespace
\textbf{Average (27 labels)} & \textbf{0.23 [0.16, 0.31]} & \textbf{0.23 [0.17, 0.33]} & 0.45 [0.36, 0.55] & 0.23 [0.17, 0.30] & \textbf{0.20 [0.14, 0.30]} & 0.47 [0.38, 0.57] \\
\bottomrule
\end{tabular*}
\label{tab:sca-dsc-asd-mp}
\end{threeparttable}
\end{table}

Additionally, the DSC and ASD values were computed against \go{} for the MPM-MTw, MPM-PDw and MPM-\tow{} contrasts and are reported in Tables \ref{tab:sca-dsc-asd-mtw}, \ref{tab:sca-dsc-asd-pdw} and \ref{tab:sca-dsc-asd-t1w} respectively in Appendix \ref{appB}. The same trend was observed, with \go{ 2.0-n2} consistently achieving the best DSC and ASD values overall, followed closely by \go{ 2.0-n1}, while \sys{} was outperformed. Ultimately, both variants of \go{ 2.0} demonstrated highly comparable performance with the original \go{} for the brain segmentation task.

Ultimately, the \go{ 2.0-n2} variant was selected as the optimal variant, as it (1) resolved the limitation related to enlarged ventricles, (2) provided the most accurate delineation of the cerebellum WM and cortex, and (3) achieved all of this while preserving the overall high performance of the original \go{}. Consequently, the \go{ 2.0-n1} variant was disregarded for the remaining analyses.

\subsection{\sca{-NPC}: Assessment of segmentation methods in SCA subjects}

\begin{figure}[htbp]
    \begin{center}
        \includegraphics[width=0.96\textwidth]{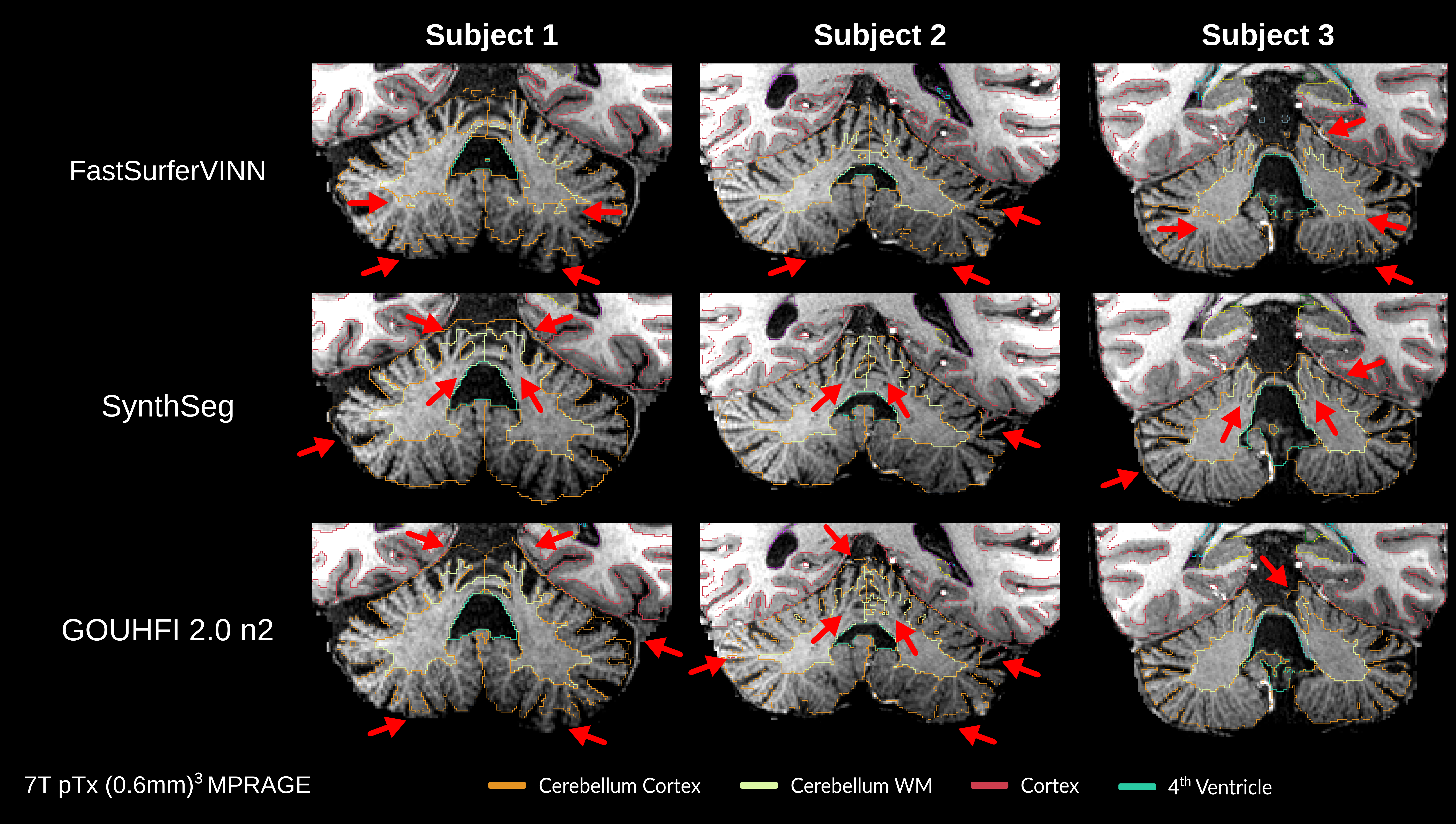}
        \caption{Brain segmentation results produced by \fsv{} (top row), \sys{} (middle row) and \go{ 2.0-n2} (bottom row) for SCA subjects overlaid on a zoomed-in coronal view of the \mpr{} used as input (not N4-corrected). All images and segmentations have a resolution of \sixi{}. Red arrows show segmentation irregularities and errors.}
        \label{fig:sca-npc-mprage}
    \end{center}
\end{figure}

While the previous results section used healthy subjects at 7T, the performance of \go{ 2.0-n2} against \fsv{} and \sys{} for three SCA subtypes are shown in Figure \ref{fig:sca-npc-mprage} with a special focus on the cerebellum. All techniques struggled to accurately segment the cerebellar WM and cortex for the three SCA subjects. While \fsv{} had the best overall delineation of the cortex-CSF boundary, it systematically missed some posterior sections of the cerebellum for all subjects. Moreover, posterior WM seemed particularly challenging to identify for \fsv{} compared to the two other techniques. Conversely, \sys{} and \go{ 2.0-n2} over-segmented the anterior cerebellar WM, with \sys{} exhibiting a more pronounced error. Overall, \sys{} successfully identified all cerebellar cortex regions, but consistently over-segmented them across all test cases. For the second subject, \go{ 2.0-n2} mislabeled some brain cortex voxels as cerebellum cortex. 

\subsection{TIV estimation}

\begin{figure}[htbp]
    \begin{center}
        \includegraphics[width=0.9\textwidth]{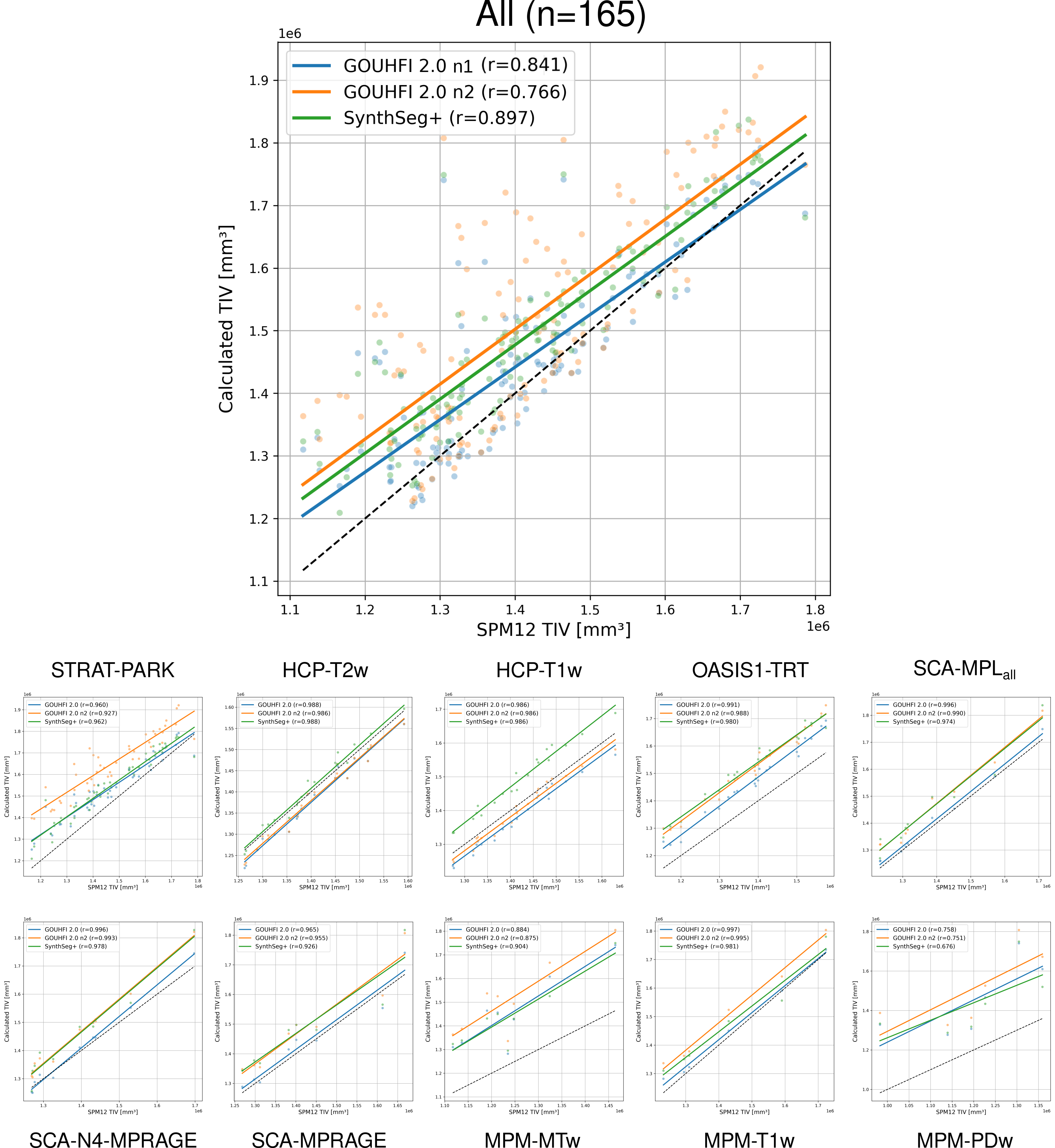}
        \caption{Scatter plots of the TIV values computed by \go{ 2.0-n1} (blue), \go{ 2.0-n2} (orange) and \sys{} (green) versus \spm{}. The black dashed line corresponds to the identity line whereas solid lines correspond to the linear regression of all cases for each technique. The Pearson's correlation coefficient (r) is shown in the legend. On top, all datasets are combined into one plot with the scatter plots for all individual datasets below.}
        \label{fig:tiv}
    \end{center}
\end{figure}

The TIV estimations from \go{ 2.0-n1}, \go{ 2.0-n2} and \sys{} compared to the reference \spm{} are shown in Figure \ref{fig:tiv}. Overall, all three techniques generated TIV values larger than \spm{} except for \go{ 2.0-n1} which tended to underestimate at large TIV values. All techniques had a Pearson's correlation coefficient (r) above 0.75, demonstrating a strong positive linear correlation with \spm{}, with \go{ 2.0-n2} being the lowest at 0.766 and \sys{} the highest at 0.897. On the dataset-specific level, all techniques calculated TIV values larger than \spm{} expect for both contrasts for the HCP. For the HCP, both versions of \go{ 2.0} gave values slightly smaller than \spm{} whereas \sys{} produced larger values, especially for the HCP-\tow{}. The MPM-PDw and MPM-MTw had substantially larger TIV estimates than \spm{} and were the only datasets with r $<$0.9. 

\subsection{7T Parkinson’s Disease volumetry: Impact of an expanded training dataset}

\begin{figure}[htbp]
    \begin{center}
        \includegraphics[width=0.92\textwidth]{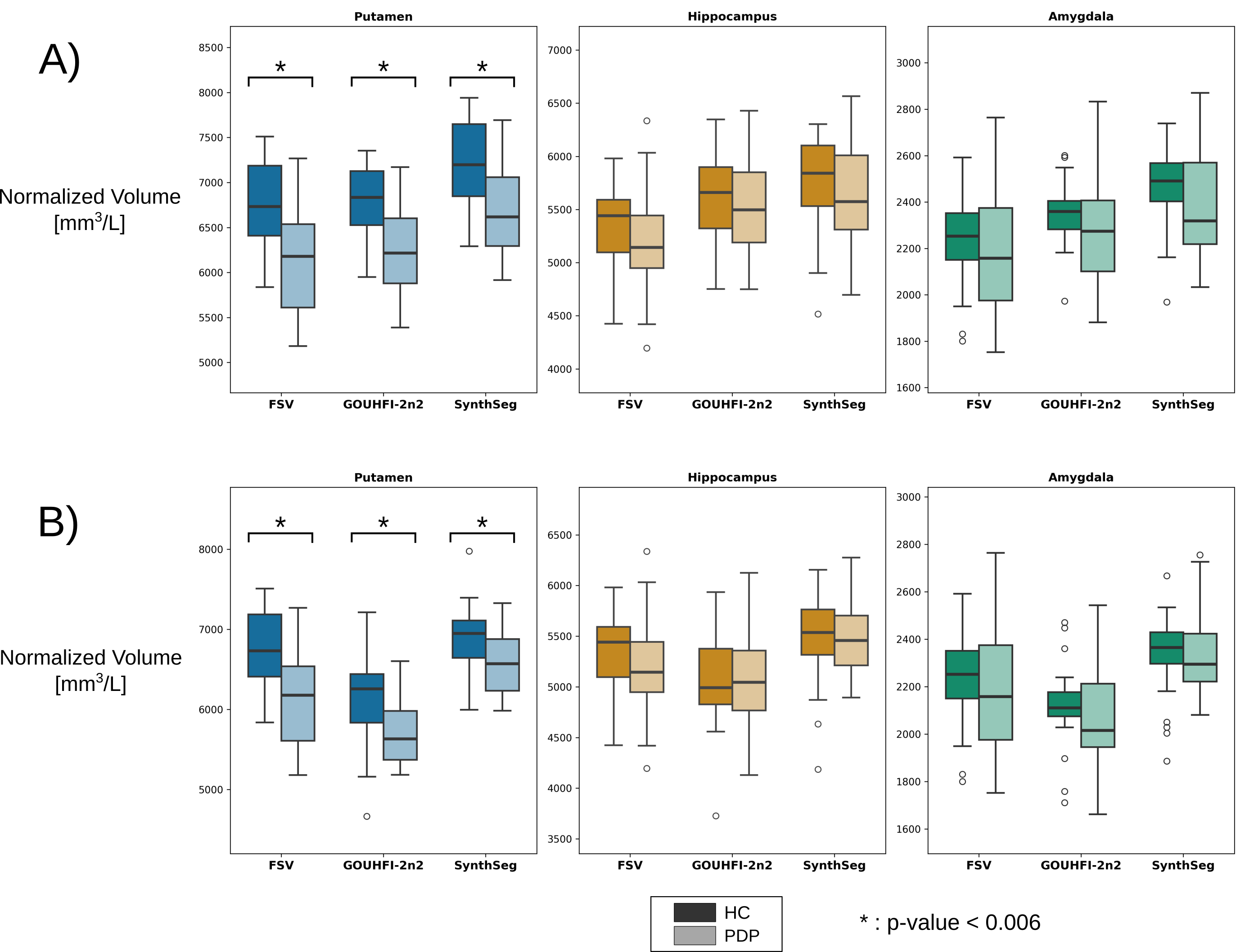}
        \caption{Box plots showing the normalized volumes computed by \fsv{} (left), \go{ 2.0-n2} (middle) and \sys{} (right) for healthy controls (HC) and people with Parkinson's disease (PwP) for the putamen, hippocampus and amygdala. In A), \spm{} was used for estimating the TIV for normalizing the volumes. In B), the same results are shown but \go{ 2.0-n2} and \sys{} were used to estimate the TIV for each respectively (\spm{} was used for \fsv{}). For putamen, the three techniques had a statistically significant difference in volume between HC and PwP after Bonferroni correction for both A) and B).}
        \label{fig:boxp-volum}
    \end{center}
\end{figure}

The results from the same volumetric analysis as done in \cite{fortin2025gouhfi} are shown in Figure \ref{fig:boxp-volum} A). Overall, the same consistent decrease trend between HC and PwP as observed for the original \go{} for the putamen, hippocampus and amygdala was observed for \go{ 2.0-n2}. It was only for putamen that all techniques presented a statistically significant difference between both HC and PwP sub-groups. For \sys{}, the median volumes measured were systematically larger than the rest. The p-values calculated are reported in Table \ref{tab:roi-pvalues}. In addition, since TIV estimation is part of the \go{ 2.0} toolbox, an independent volumetry analysis with normalized volume values were performed without the use of \spm{}. The same volumetric analysis, using TIV estimates from \go{2.0-n2} and \sys{}, is shown in Figure \ref{fig:boxp-volum} B), with corresponding p-values reported in Table \ref{tab:roi-pvalues}. Albeit smaller than the normalized volumes reported using \spm{}, the new normalized volumes resulted in the same statistical significance for the putamen. For both \fsv{} and the original \go{}, the TIV was not calculated independently due to differing definitions for the CSF label. An additional figure showing the volumetry results for the original \go{}, \go{ 2.0-n2} with TIV from \spm{} and \go{ 2.0-n2} with independent TIVs is provided in Appendix \ref{appC}.

\begin{table}[h]
\centering
\begin{threeparttable}
\caption{P-values computed for the putamen, hippocampus and amygdala for the different segmentation methods and TIV calculation approaches tested in this work. The Bonferroni-correct significance threshold p-value was set at $<$ 0.006. P-values with $^*$ are statistically significant.}
\footnotesize
\setlength\tabcolsep{0pt}
\begin{tabular*}{0.85\linewidth}{@{\extracolsep{\fill}} lrrrr}
\toprule
\textbf{TIV: \spm{}} & \fsv{} & \go{} & \go{ 2.0-n2} & \sys{} \\
\midrule
Putamen      &  0.002$^*$ & 0.004$^*$ & 0.001$^*$ & 0.0004$^*$ \\
Hippocampus  &  0.22 & 1.0 & 0.50 &  0.31\\
Amygdala     &  0.36 & 0.06 & 0.09 & 0.27 \\
\toprule
\textbf{TIV: Independent} & \fsv{} & \go{} & \go{ 2.0-n2} & \sys{} \\
\midrule
Putamen      & - & - & 0.004$^*$ & 0.003$^*$  \\
Hippocampus  & - & - & 0.92 & 0.83 \\
Amygdala     & - & - & 0.29 & 0.64 \\
\bottomrule
\end{tabular*}
\label{tab:roi-pvalues}
\end{threeparttable}
\end{table}

\subsection{\go{ 2.0}: Cortex parcellation performance}

The quantitative assessment of \go{ 2.0's} cortical parcellation against manual cortical delineations are reported in Table \ref{tab:ctx-man}. \go{ 2.0} ranked second behind \fsv{} regarding median DSC and ASD values while being 6 DSC points higher than \sys{} and \ants{}. 

\begin{table}[h]
\centering
    \begin{threeparttable}
    \caption{Median DSC and ASD values (with IQR) of the cortex labels segmented by \fsv{}, \go{ 2.0}, \sys{} and \ants{} compared to the manual delineations from the MindBoggle101 dataset (n=20). 52 (out of 62) labels were included in the calculations since \sys{} follows a different labeling convention than the three others. The cortex labels produced by \go{ 2.0-n2} were selected as inputs to the cortex parcellation model. The highest DSC (and lowest ASD) values among the three methods are shown in bold.}
        \footnotesize
        \setlength\tabcolsep{0pt}
        \begin{tabular*}{0.95\linewidth}{@{\extracolsep{\fill}} lrrrr}
            \toprule
             & \fsv{} & \go{ 2.0} & \sys{} & \ants{} \\
            \midrule
            DSC & \textbf{0.86 [0.81, 0.88]} & 0.79 [0.75, 0.83] & 0.73 [0.66, 0.78] & 0.73 [0.68, 0.75] \\[0.25em]
            ASD [mm] & \textbf{0.32 [0.22, 0.54]} & 0.50 [0.38, 0.71] & 0.70 [0.51, 1.12] & 0.64 [0.54, 0.86] \\
            \bottomrule
        \end{tabular*}
        \label{tab:ctx-man}
    \end{threeparttable}
\end{table}

Furthermore, the cortex parcellations from \go{ 2.0} for the HCP-\tow{} dataset were compared to \sys{} and \ants{}, using \fsv{} as the reference this time. The median DSC and ASD values are listed in Table \ref{tab:ctx-hcp}. \go{ 2.0} outperformed \sys{} and \ants{} by 11 and 16 DSC points, respectively, across all test subjects. 

\begin{table}[h]
\centering
    \begin{threeparttable}
    \caption{Median DSC and ASD values (with IQR) for all cortex labels segmented by \go{ 2.0}, \sys{} and \ants{} compared to the cortex parcellation produced by \fsv{} from the HCP-YA test dataset (n=20). 52 (out of 62) labels were included in the calculations since \sys{} follows a different labeling convention than the three others. The highest DSC (and lowest ASD) values among the three methods are shown in bold.}
        \small
        \setlength\tabcolsep{0pt}
        \begin{tabular*}{0.75\linewidth}{@{\extracolsep{\fill}} lrrr}
            \toprule
             & \go{ 2.0} & \sys{} & \ants{} \\
            \midrule
            DSC & \textbf{0.88 [0.85, 0.90]} & 0.77 [0.72, 0.81] & 0.72 [0.68, 0.75] \\[0.25em]
            ASD [mm] & \textbf{0.29 [0.25, 0.35]} & 0.56 [0.45, 0.85] & 0.66 [0.55, 0.80] \\
            \bottomrule
        \end{tabular*}
        \label{tab:ctx-hcp}
    \end{threeparttable}
\end{table}

\section{Discussion}\label{disc}

In this study, we are proposing \go{ 2.0}, a segmentation toolbox capable of segmenting the brain and cortex into 35 and 62 labels, respectively, from MR images regardless of their contrast, resolution or field strength. \go{ 2.0} maintained highly comparable performance to its predecessor for brain segmentation and volumetry, while extending its functionality through cortical parcellation where it out-performed \sys{} and \ants{}. Notably, it also demonstrated improved robustness in non-healthy anatomies with enlarged ventricles, something that the original method showed limitations for. For SCA subjects, \go{ 2.0} demonstrated certain shortcomings for the cerebellum segmentation, though both \fsv{} and \sys{} also revealed notable weaknesses.

\subsection{\go{ 2.0}: Addressing challenges in older and clinical cohorts} 

While testing the original \go{} \citep{fortin2025gouhfi}, incidental observations of the limitations highlighted in Figure \ref{fig:why-2p0} prompted the idea of enhancing the diversity of the training dataset. One known limitation of the original training dataset was the absence of aged and/or non-healthy brain anatomies. The original \go{} exhibited limitations not observed for \fsv{} or \sys{}, likely due to the absence of non-healthy anatomies in its training dataset, which were included in the latter two. In order to address this issue, 32 aged and demented subjects were incorporated based on an ad hoc visual assessment of subjects from the OASIS1 dataset and only selecting subjects exhibiting substantially enlarged ventricles. Ultimately, segmentation errors like the ones observed in the original \go{} were completely resolved when qualitatively assessing \go{ 2.0's} output segmentations for the same test subjects, while also being absent in all other test subjects assessed.

Additionally, this limited capacity to delineate enlarged ventricles also highlighted a limitation of the generative model, which is its incapacity to generate synthetic images with anatomical features observed in non-healthy brains. Until further developments on the DR approach for brain segmentation are proposed, using an anatomically-varied training label maps is highly recommended. As outlined in \cite{hoffmann2025domain}, recent developments in diffusion and flow-based DL models show high potential for improving the image synthesis models, hopefully contributing to a more robust and varied synthesizing process.

\subsection{\sca{}: Comparing \go{'s} variants in healthy controls at 7T}\label{disc:sca}

Figure \ref{fig:why-2p0n2} presented the unexpected weakness of \go{ 2.0-n1} struggling with proper identification of posterior cerebellar cortex, exclusively observed for that model. Surprisingly, this suboptimal cerebellar cortex segmentation for \go{ 2.0-n1} did not seem to arise from typical inhomogeneities observed at UHF, since this was mostly observed in homogeneous N4-corrected pTx images (as in Figure \ref{fig:why-2p0n2}) while not being observed in inhomogeneous 1 Tx 7T images. Ultimately, one possible explanation for this unexpected behavior for \go{ 2.0-n1} could be the randomness involvement in (1) the generation of synthetic images and (2) the initial seeding of the DL training process as shown in \cite{aakesson2024random}.

The superior identification of cerebellar WM branches by \go{ 2.0-n2} as shown for both subjects in Figure \ref{fig:why-2p0n2} makes it a tool of choice for researchers working with the cerebellum, outperforming \fsv{} and \sys{}. Improvements are also evident when comparing the original \go{} to the latest two variants of \go{ 2.0}. As discussed in \cite{fortin2025gouhfi}, even though \sys{} also uses a DR approach, the difference in training resolutions (\onei{} vs \sevi{}) seems especially important for detecting small regions such as thin cerebellar WM branches.

\subsection{\sca{-NPC}: Assessment of segmentation methods in SCA subjects}

Overall, no segmentation technique demonstrated a clear superiority over the others. It is important to mention that this section was focused on the cerebellum only. First, \fsv{} performed relatively well considering that these test cases were outside its optimized domain (i.e., \sixi{} 7T images). It was the highest performing technique to properly delineate regions with substantial cortex atrophy. This improved delineation in atrophied cortical regions was most pronounced in the anterior lobe of the cerebellum where both \sys{} and \go{ 2.0-n2} failed at accurately identifying the CSF-cortex border (cf. results for subject 1). For subjects 2 and 3 with less atrophy, this was less pronounced. This result can be surprising considering that both DR approaches were trained using perfectly aligned training image and label maps from the synthetic images produced by the generative model from the label maps. In other words, one could have expected a better identification of the CSF-GM border by \sys{}/\go{ 2.0} since the two regions have high contrast between each other. In the end, this circles back to the previous section discussing that a variation of anatomies inside the training corpus is still important, even when using synthetic training images. However, for subject 3, \go{ 2.0-n2} demonstrated a great capacity at accurately identifying the CSF-cortex border, except for the anterior lobe. 

On the other hand, \fsv{} considerably underestimated the inferior region of the posterior lobe of the cerebellar WM in two of the three test subjects while also systematically failing at properly delineating the inferior posterior cerebellar cortex, possibly due to the reduced signal due to inhomogeneities. Both issues can be problematic in volumetric studies where under-estimations of both WM and cortex volumes due to segmentation errors can bias the results. For \sys{}, the users should be aware of the opposite trend: over-estimation of WM and cortex volumes. As discussed in the last paragraph, \sys{} wrongly segmented a substantial amount of cortex and CSF as cerebellum WM. For the cortex, \sys{} performed the worst at identifying small atrophy regions, resulting in the cortex label to be segmented as a oversimplified continuous shape instead of following the "branchy" nature of the atrophied cerebellum cortex for SCA subjects. For \go{2.0-n2}, the marginal WM overestimation and improved cortex–CSF border identification, compared to \sys{}, indicate that it is a better alternative for high-resolution UHF images.

Regarding the mislabeling by \go{ 2.0} of the cerebellar cortex with cerebral WM (cf. Subject 2 in Figure \ref{fig:sca-npc-mprage}), a pathological hallmark of SCAs is the degeneration of the so-called \textit{Purkinje-cell layer}, leading to a gliotic aspect of affected cerebellar areas and thereby a thinning of the cerebellar cortex. The anterior lobe of the cerebellum is known to be particularly affected in the most frequent types of SCA in central and northern Europe presented here, sometimes even before the onset of motor symptoms. These changes are known to be visible in structural MRI and are used diagnostically by clinicians. It is therefore noteworthy that we observed particular difficulties of \go{ 2.0} to segment the cerebellar cortex of the anterior lobe. This might indicate that the tool's inability to identify cortical areas of the cerebellum here, points to the tool accidentally picking up an important pathophysiological feature of the disease\cite{Faber2021-ep,vieira2025cerebellum,seidel2012brain}.

In the end, assessing which technique is the most accurate is quite challenging and should be done on a task and dataset basis. Thus, we recommend researchers to test, if possible, more than one technique and do a visual assessment of the results before selecting the segmentation technique to be used for their analyses.

\subsection{TIV estimation}

One interesting result from the TIV estimations was that the only cases where the \go{ 2.0} variants under-estimated the TIV, compared to \spm{}, are the two HCP contrasts. In contrast, \sys{} exhibited a systematic overestimation of the TIV for the HCP dataset and, more broadly, across all ten test datasets. One difference compared to all other test datasets was that the brain extraction step was not performed as part of this work, since it was already performed as part of the HCP preprocessed data. In both cases, the brain extraction process was very accurate, with no inclusion of extra tissue. For STRAT-PARK, only brain-extracted images were provided, with Nighres \citep{huntenburg2018nighres} used for the brain extraction step. For most subjects in that dataset, a substantial amount of extra-cerebral tissue was still present in the brain-extracted images as shown in Figure \ref{fig:why-2p0}. For the remaining datasets, the brain extraction step was performed with the \ants{} brain extraction tool implemented as part of the \go{} toolbox. The results were visually assessed, and cases with minor extra-cerebral tissue were still retained (negligible compared to START-PARK), as this had no impact on brain segmentation aside from the CSF label. Indeed, for \go{2.0-n2}, visual inspection revealed that the CSF label often incorporated extra-cerebral tissue when brain extraction quality was low. On the other hand, \go{ 2.0-n1} was not as affected by extra-cerebral tissue, resulting in less TIV overestimation. While the \go{ 2.0-n1} variant resulted in higher-quality TIV estimates than \go{ 2.0-n2}, the latter was still selected as the superior variant due to the reasons stated in section \ref{disc:sca}.

Conclusively, if optimal TIV estimates from \go{ 2.0-n2} are of relevance for your work, high-quality brain extraction of the input images is recommended to mitigate the observed overestimation of TIV values.

\subsection{7T Parkinson’s Disease volumetry: Impact of an expanded training dataset}

Even with improved delineation of the hippocampus with \go{ 2.0-n2}, the volumetry analysis conclusions remained unchanged from the original \go{}. The original volumetry results are shown in Appendix \ref{appC}. This is particularly relevant for the volumetry analysis conducted without using \spm{} for TIV values (Figure \ref{fig:boxp-volum} B)). While the normalized volumes are smaller than the ones computed with \spm{}, this is not surprising considering the known over-estimation of the TIV, as discussed in the previous section. Ultimately, avoiding the use of \spm{} considerably simplifies and streamlines the volumetric analysis from a technical standpoint.

While \fsv{} was not considered the reference for earlier brain segmentation quantitative comparison in this study, it was selected as the reference for the volumetry analysis. Based on extensive visual inspection of the segmentation results, \fsv{} was deemed as the technique with the most accurate delineation of the hippocampus-ventricle laterally. However, in subjects with strong signal inhomogeneities, medial portions of both the hippocampus and amygdala may be excluded, causing lower estimated volumes and possibly explains the lower estimates in Figure \ref{fig:boxp-volum} A) compared to \go{ 2.0-n2} and \sys{}.

\subsection{\go{ 2.0}: Cortex parcellation performance}

\go{ 2.0} outperformed \sys{} and \ants{} for both comparisons against manual delineations for OASIS1, and against \fsv{} for HCP-\tow{}, as presented with Tables \ref{tab:ctx-man} and \ref{tab:ctx-hcp}. Interestingly, \go{ 2.0} was the only technique without subjects from the OASIS1 dataset as part of its training dataset. For the HCP-\tow{}, only \ants{} did not include HCP subjects inside its training dataset.

It is important to reiterate that the cortex parcellation task as part of \go{ 2.0} is executed by a different U-Net than the brain segmentation, using solely the cortex label as input. This comes with the advantage that any cortex segmentation can be used as the input for the cortex parcellation task, no matter its origin. By default, \go{ 2.0} will extract and use the cortex label segmented by the brain segmentation task done by \go{ 2.0-n2} as input. Conversely, the parcellation quality may be compromised when the input cortex segmentation is suboptimal. Overall, these initial findings position \go{ 2.0} as a quite attractive and versatile solution for cortex parcellation, particularly at UHF.
  
\subsection{Limitations}

The main limitation of the \go{ 2.0} toolbox is the brain extraction requirement. As already discussed in the original publication of \go{}, most segmentation techniques do not share this requirement and is partly explained by lack of robust whole-head segmentation techniques for UHF images \citep{fortin2025gouhfi}. Future advances in whole-head segmentation algorithms for UHF images with low signal outside the head, or the use of generative diffusion models to generate synthetic head labels, could help address this issue.

Moreover, in line with the previous limitation, the precision of the brain extraction procedure can have a direct effect on TIV estimates generated by \go{ 2.0}. Users should be aware of possible TIV over-estimations from \go{ 2.0-n2} when non-negligible extra-cerebral tissue remains in the images. Future developments on \go{ 2.0} should focus on enabling a more accurate TIV computation pipeline.

\section{Conclusions}\label{conc}

In summary, we propose \go{ 2.0}, a new and improved implementation of the original \go{} segmentation toolbox. While preserving the contrast- and resolution-agnostic properties and robustness to UHF signal inhomogeneities of the original implementation, \go{ 2.0} introduces additional features, including cortical parcellation and a volumetry pipeline for extracting both regional volume and TIV values. More precisely, \go{ 2.0} has now two independently trained segmentation tasks using state-of-the-art 3D U-Nets. The first U-Net enables the segmentation of brain images of any contrast, resolution or field strength into 35 labels, with improved performance on aged and non-healthy clinical cohorts compared to the original \go{}. The second U-Net allows for the parcellation of the cortex into 62 labels, being the first DL method enabling robust cortex parcellation at UHF. Ultimately, future work will aim to overcome the brain extraction limitation of \go{ 2.0}, thereby enhancing both its usability and TIV estimation accuracy.

\section*{Data and Code Availability}

The source code for \go{ 2.0} is available on GitHub at \href{https://github.com/mafortin/GOUHFI}{https://github.com/mafortin/GOUHFI}.

All MRI datasets used in this article are open source repositories freely available through the links provided in footnotes of section \ref{meth} except for the two \sca{} and STRAT-PARK datasets. They are not publicly available due to data protection regulations.

\section*{Author Contributions}

\textbf{MAF}: Conceptualization, Data Curation, Methodology, Formal analysis, Investigation, Software, Validation, Writing-original draft, Writing-review \& editing, Visualization. \textbf{ALK}: Writing-review \& editing, Data Curation, Validation, Formal analysis. \textbf{KES}: Writing-review \& editing, Data acquisition. \textbf{NK}: Writing-review \& editing, Data acquisition. \textbf{CT}: Writing-review \& editing, Funding acquisition. \textbf{PEG}: Supervision, Resources, Project administration, Writing-review \& editing, Funding acquisition.

\section*{Funding}

This project and \sca{} data acquisition were supported through the following funding organizations under the aegis of JPND: Belgium, The Fund for Scientific Research (F.R.S.-FNRS; funding code PINT-MULTI/BEJR.8006.20); Germany, Federal Ministry of Education and Research (BMBF; funding codes 01ED2109A/B); and Norway, The Research Council of Norway (RCN; funding code 322980).

\section*{Declaration of Competing Interests}

The authors do not declare any competing interests.

\section*{Acknowledgments}

First, we would like to thank the SCAIFIELD consortium (Principal Investigator: Tony Stöcker, Data acquisition \sca{-NPC} dataset: Nicolas Kunath) and the STRAT-PARK study (Study directors: Charalampos Tzoulis \& Mandar Jog) for supporting the data acquisition and funding related to each dataset used in this article. Moreover, we would like to thank Michael Staff Larsen for the constructive feedback and discussions about \go{ 2.0}. \\
We would also like to thank all the publicly available datasets used in this work. First, data provided as part as the Human Connectome Project, WU-Minn Consortium (Principal Investigators: David Van Essen and Kamil Ugurbil; 1U54MH091657) funded by the 16 NIH Institutes and Centers that support the NIH Blueprint for Neuroscience Research; and by the McDonnell Center for Systems Neuroscience at Washington University were used in this publication. Moreover, data provided in part by the OASIS-2: Longitudinal: Principal Investigators: D. Marcus, R, Buckner, J. Csernansky, J. Morris; P50 AG05681, P01 AG03991, P01 AG026276, R01 AG021910, P20 MH071616, U24 RR021382 was also used.\\ 
Finally, we would also like to thank individually all these following projects/initiatives for sharing their dataset and making this article possible: (1) the UltraCortex: Submillimeter Ultra-High Field 9.4 T1 Brain MR Image Collection and Manual Cortical Segmentations dataset (DOI:\href{https://doi.org/10.18112/openneuro.ds005216.v1.1.0}{https://doi.org/10.18112/open\\neuro.ds005216.v1.1.0}), (2) Autism Brain Imaging Data Exchange II (ABIDE-II) initiative (DOI: \href{https://dx.doi.org/10.21227/y3v9-b041}{https://\\dx.doi.org/10.21227/y3v9-b041}) and (3) MindBoggle101 (DOI:\href{https://doi.org/10.3389/fnins.2012.00171}{https://doi.org/10.3389/fnins.2012.00171}). 


\printbibliography

\section{Appendix}

\subsection{DSC and ASD tables for MPM-MTw, MPM-PDw and MPM-\tow{} images for the \sca{} dataset.}
\label{appB}

\begin{table}[h]
\centering
\begin{threeparttable}
\caption{Median DSC and ASD values (with IQR) computed for the MPM-MTw contrast in the SCAIFIELD dataset (n=10) using \go{ 2.0-n1}, \go{ 2.0-n2} and \sys{}. The ground truth is the segmentation created by \go{}. The highest DSC (and lowest ASD) values among the three methods for each structure are shown in bold.}
\scriptsize
\begin{tabular*}{0.75\linewidth}{@{\extracolsep{\fill}}l ccc}
\toprule
\textbf{DSC} & \multicolumn{3}{c}{MPM-MTw} \\
            & \go{ 2.0-n1} & \go{ 2.0-n2} & \sys{} \\
\midrule
WM                & \textbf{0.97 [0.97, 0.98]} & \textbf{0.97 [0.97, 0.98]} & 0.92 [0.92, 0.93] \\
\addlinespace
Cortex            & \textbf{0.94 [0.93, 0.94]} & \textbf{0.94 [0.94, 0.94]} & 0.87 [0.87, 0.88] \\
\addlinespace
Putamen           & \textbf{0.97 [0.97, 0.97]} & \textbf{0.97 [0.97, 0.97]} & 0.91 [0.91, 0.92]\\
\addlinespace
Thalamus          & \textbf{0.96 [0.96, 0.96]} & \textbf{0.96 [0.94, 0.96]} & 0.93 [0.92, 0.93] \\
\addlinespace
Pallidum          & 0.93 [0.92, 0.94] & \textbf{0.94 [0.93, 0.94]} & 0.85 [0.81, 0.86] \\
\addlinespace
Cerebellum WM     & 0.94 [0.94, 0.95] & \textbf{0.95 [0.95, 0.95]} & 0.89 [0.89, 0.90] \\
\addlinespace
Cerebellum Cortex & 0.95 [0.93, 0.95] & \textbf{0.96 [0.95, 0.96]} & 0.93 [0.93, 0.93] \\
\addlinespace
\textbf{Average (27 labels)} & \textbf{0.94 [0.92, 0.96]} & \textbf{0.94 [0.92, 0.96]} & 0.89 [0.85, 0.92] \\
\bottomrule
\end{tabular*}
\vspace{1em}
\begin{tabular*}{\linewidth}{@{\extracolsep{\fill}}l ccc}
\toprule
\textbf{ASD [mm]} & \multicolumn{3}{c}{MPM-MTw} \\
                  & \go{ 2.0-n1} & \go{ 2.0-n2} & \sys{} \\
\midrule
WM                & \textbf{0.12 [0.11, 0.12]} & 0.13 [0.13, 0.14] & 0.33 [0.33, 0.34] \\
\addlinespace
Cortex            & \textbf{0.22 [0.21, 0.24]} & 0.24 [0.23, 0.26] & 0.42 [0.41, 0.43] \\
\addlinespace
Putamen           & \textbf{0.16 [0.16, 0.17]} & 0.17 [0.16, 0.18] & 0.51 [0.48, 0.52] \\
\addlinespace
Thalamus          & \textbf{0.28 [0.27, 0.31]} & 0.30 [0.27, 0.36] & 0.49 [0.47, 0.54] \\
\addlinespace
Pallidum          & 0.31 [0.28, 0.33] & \textbf{0.28 [0.26, 0.31]} & 0.59 [0.55, 0.73] \\
\addlinespace
Cerebellum WM     & \textbf{0.20 [0.18, 0.21]} & \textbf{0.20 [0.18, 0.23]} & 0.42 [0.40, 0.45] \\
\addlinespace
Cerebellum Cortex & 0.39 [0.37, 0.47] & \textbf{0.34 [0.31, 0.42]} & 0.51 [0.49, 0.55] \\
\addlinespace
\textbf{Average (27 labels)} & 0.27 [0.18, 0.37] & \textbf{0.26 [0.19, 0.34]} &  0.48 [0.41, 0.58] \\
\bottomrule
\end{tabular*}
\label{tab:sca-dsc-asd-mtw}
\end{threeparttable}
\end{table}

\begin{table}[h]
\centering
\begin{threeparttable}
\caption{Median DSC and ASD values (with IQR) computed for the MPM-PDw contrast in the SCAIFIELD dataset (n=10) using \go{ 2.0-n1}, \go{ 2.0-n2} and \sys{}. The ground truth is the segmentation created by \go{}. The highest DSC (and lowest ASD) values among the three methods for each structure are shown in bold.}
\scriptsize
\begin{tabular*}{0.75\linewidth}{@{\extracolsep{\fill}}l ccc}
\toprule
\textbf{DSC} & \multicolumn{3}{c}{MPM-PDw} \\
            & \go{ 2.0-n1} & \go{ 2.0-n2} & \sys{} \\
\midrule
WM                & 0.95 [0.95, 0.96] & \textbf{0.96 [0.95, 0.96]} & 0.90 [0.89, 0.91] \\
\addlinespace
Cortex            & \textbf{0.92 [0.91, 0.92]} & \textbf{0.92 [0.91, 0.92]} & 0.85 [0.84, 0.86]  \\
\addlinespace
Putamen           & 0.94 [0.93, 0.95] & \textbf{0.95 [0.93, 0.96]} & 0.87 [0.85, 0.89] \\
\addlinespace
Thalamus          & \textbf{0.94 [0.93, 0.94]} & 0.93 [0.92, 0.94] & 0.92 [0.90, 0.92] \\
\addlinespace
Pallidum          & 0.90 [0.89, 0.92] & \textbf{0.91 [0.90, 0.92]} & 0.78 [0.74, 0.83] \\
\addlinespace
Cerebellum WM     & 0.93 [0.93, 0.94] & \textbf{0.95 [0.94, 0.95]} & 0.90 [0.89, 0.90] \\
\addlinespace
Cerebellum Cortex & 0.94 [0.93, 0.94]  & \textbf{0.95 [0.95, 0.96]} & 0.93 [0.93, 0.93] \\
\addlinespace
\textbf{Average (27 labels)} & \textbf{0.91 [0.85, 0.94]} & \textbf{0.91 [0.86, 0.94]} & 0.86 [0.80, 0.90] \\
\bottomrule
\end{tabular*}
\vspace{1em}
\begin{tabular*}{\linewidth}{@{\extracolsep{\fill}}l ccc}
\toprule
\textbf{ASD [mm]} & \multicolumn{3}{c}{MPM-PDw} \\
                  & \go{ 2.0-n1} & \go{ 2.0-n2} & \sys{} \\
\midrule
WM                & \textbf{0.20 [0.19, 0.22]} & 0.23 [0.21, 0.24] & 0.41 [0.39, 0.43] \\
\addlinespace
Cortex            & \textbf{0.33 [0.32, 0.36]} & 0.37 [0.36, 0.42] & 0.54 [0.53, 0.55] \\
\addlinespace
Putamen           & 0.31 [0.28, 0.33] & \textbf{0.28 [0.25, 0.36]} & 0.74 [0.62, 0.86] \\
\addlinespace
Thalamus          & 0.46 [0.45, 0.49] & \textbf{0.44 [0.38, 0.51]} & 0.57 [0.54, 0.61] \\
\addlinespace
Pallidum          & \textbf{0.41 [0.37, 0.50]} & 0.43 [0.37, 0.48] & 0.85 [0.69, 1.02] \\
\addlinespace
Cerebellum WM     & 0.27 [0.24, 0.31] & \textbf{0.24 [0.21, 0.31]} & 0.35 [0.34, 0.39]  \\
\addlinespace
Cerebellum Cortex & 0.47 [0.45, 0.52] & \textbf{0.45 [0.41, 0.50]} &  0.52 [0.50, 0.53] \\
\addlinespace
\textbf{Average (27 labels)} & 0.43 [0.33, 0.58] & \textbf{0.42 [0.32, 0.56]} & 0.61 [0.52, 0.88] \\
\bottomrule
\end{tabular*}
\label{tab:sca-dsc-asd-pdw}
\end{threeparttable}
\end{table}

\begin{table}[h]
\centering
\begin{threeparttable}
\caption{Median DSC and ASD values (with IQR) computed for the MPM-\tow{} contrast in the SCAIFIELD dataset (n=10) using \go{ 2.0-n1}, \go{ 2.0-n2} and \sys{}. The ground truth is the segmentation created by \go{}. The highest DSC (and lowest ASD) values among the three methods for each structure are shown in bold.}
\scriptsize

\begin{tabular*}{0.75\linewidth}{@{\extracolsep{\fill}}l ccc}
\toprule
\textbf{DSC} & \multicolumn{3}{c}{MPM-\tow{}} \\
            & \go{ 2.0-n1} & \go{ 2.0-n2} & \sys{} \\
\midrule
WM                & \textbf{0.97 [0.96, 0.97]} & \textbf{0.97 [0.97, 0.97]} & 0.90 [0.90, 0.91] \\
\addlinespace
Cortex            & \textbf{0.95 [0.94, 0.95]} & \textbf{0.95 [0.95, 0.95]} & 0.87 [0.85, 0.88]   \\
\addlinespace
Putamen           & 0.93 [0.93, 0.95] & \textbf{0.94 [0.94, 0.95]} & 0.88 [0.87, 0.89] \\
\addlinespace
Thalamus          & \textbf{0.96 [0.95, 0.96]} & \textbf{0.96 [0.95, 0.96]} & 0.93 [0.92, 0.94] \\
\addlinespace
Pallidum          & 0.87 [0.84, 0.89] & \textbf{0.88 [0.85, 0.91]} & 0.80 [0.79, 0.82] \\
\addlinespace
Cerebellum WM     & \textbf{0.92 [0.90, 0.92]} & \textbf{0.92 [0.91, 0.93]} &  0.82 [0.81, 0.84] \\
\addlinespace
Cerebellum Cortex & 0.95 [0.94, 0.95] & \textbf{0.96 [0.96, 0.96]} & 0.92 [0.91, 0.92] \\
\addlinespace
\textbf{Average (27 labels)} & 0.93 [0.91, 0.95] & \textbf{0.94 [0.90, 0.96]} &  0.87 [0.82, 0.90] \\
\bottomrule
\end{tabular*}
\vspace{1em}
\begin{tabular*}{\linewidth}{@{\extracolsep{\fill}}l ccc}
\toprule
\textbf{ASD [mm]} & \multicolumn{3}{c}{MPM-\tow{}} \\
                  & \go{ 2.0-n1} & \go{ 2.0-n2} & \sys{} \\
\midrule
WM                & \textbf{0.17 [0.16, 0.18]} & \textbf{0.17 [0.16, 0.19]} & 0.42 [0.41, 0.47] \\
\addlinespace
Cortex            & \textbf{0.18 [0.17, 0.19]} & \textbf{0.18 [0.17, 0.20]} & 0.40 [0.39, 0.41] \\
\addlinespace
Putamen           & 0.34 [0.27, 0.38] & \textbf{0.31 [0.28, 0.34]} & 0.68 [0.62, 0.72] \\
\addlinespace
Thalamus          & \textbf{0.30 [0.29, 0.34]} & 0.32 [0.29, 0.35] & 0.46 [0.43, 0.51] \\
\addlinespace
Pallidum          & 0.56 [0.48, 0.67] & \textbf{0.50 [0.39, 0.62]} & 0.81 [0.63, 0.91] \\
\addlinespace
Cerebellum WM     & \textbf{0.40 [0.33, 0.45]} & 0.43 [0.35, 0.48] & 1.03 [0.93, 1.31] \\
\addlinespace
Cerebellum Cortex &  0.37 [0.36, 0.39] & \textbf{0.34 [0.31, 0.37]} & 0.62 [0.58, 0.70] \\
\addlinespace
\textbf{Average (27 labels)} & 0.32 [0.21, 0.41] & \textbf{0.31 [0.21, 0.41]} & 0.58 [0.44, 0.74] \\
\bottomrule
\end{tabular*}
\label{tab:sca-dsc-asd-t1w}
\end{threeparttable}
\end{table}

\clearpage
\subsection{Comparison of the volumetry results between the original \go{}, \go{ 2.0-n2} and \go{ 2.0-n2} without \spm12{}}
\label{appC}

\begin{figure}[h]
    \begin{center}
        \includegraphics[width=0.92\textwidth]{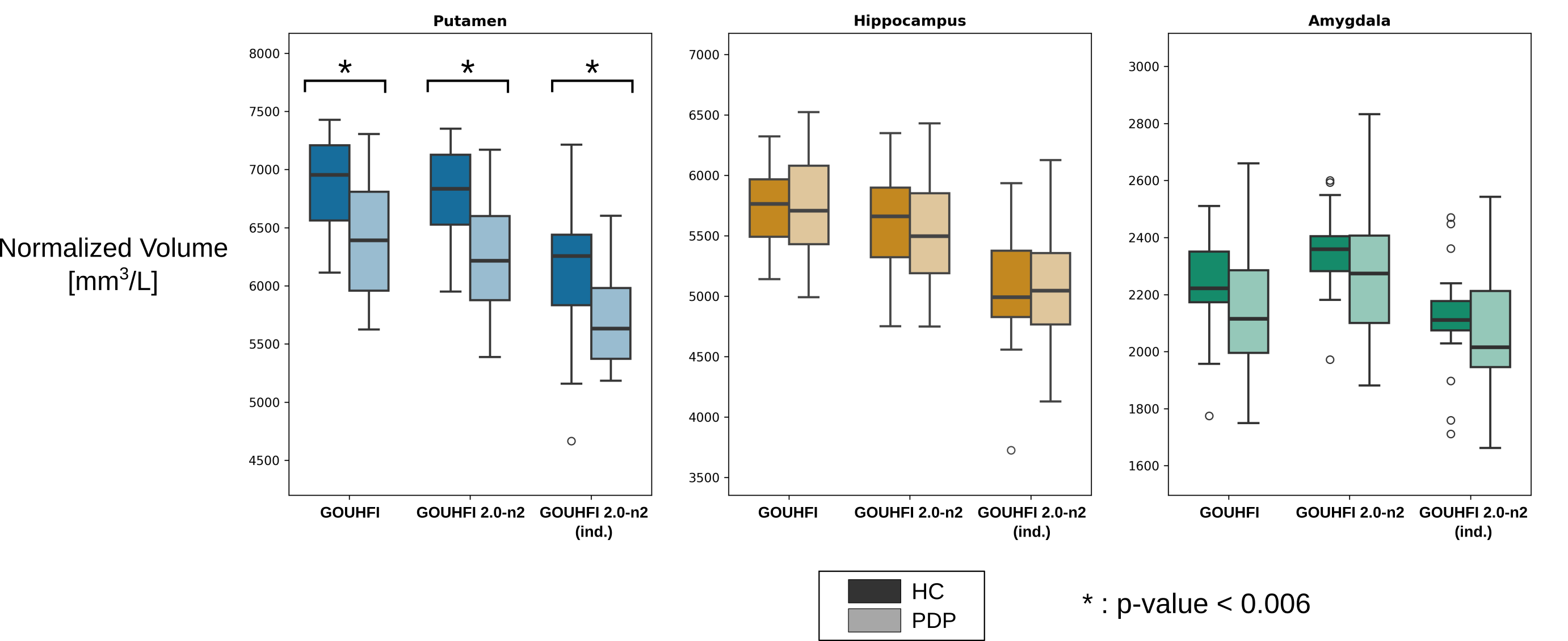}
        \caption{Box plots showing the normalized volumes measured by the original \go{} (left), \go{ 2.0-n2} with \spm{} (middle) and \go{ 2.0-n2} with independent TIV estimations (right) for healthy controls (HC) and people with Parkinson's disease (PwP) for the putamen, hippocampus and amygdala.}
        \label{fig:app-volum}
    \end{center}
\end{figure}

\end{document}